\documentclass[reprint,superscriptaddress,amsmath,amssymb,prb,]{revtex4-2}
\usepackage{graphicx} % Required for inserting images
\usepackage{amsmath}
\usepackage{indentfirst}
\usepackage{multirow}
\usepackage{subfigure}
\begin{document}

\title{Examining density wave correlations in high pressure $\rm{La_3Ni_2O_7}$ through variational Monte Carlo}
%\title{Electron correlation driven density waves in high pressure $\rm{La_3Ni_2O_7}$}
\author{Yanxin Chen}
 \affiliation{School of Physics, Peking University, Beijing 100871, People's Republic of China.}
\author{Haoxiang Chen}
 \email{hxchen@pku.edu.cn}
 \affiliation{School of Physics, Peking University, Beijing 100871, People's Republic of China.}
\author{Tonghuan Jiang}
 \affiliation{School of Physics, Peking University, Beijing 100871, People's Republic of China.}
\author{Ji Chen}
 \email{ji.chen@pku.edu.cn}
 \affiliation{School of Physics, Peking University, Beijing 100871, People's Republic of China.}
 \affiliation{Interdisciplinary Institute of Light-Element Quantum Materials and Research Center for Light-Element Advanced Materials, Peking University, Beijing 100871, People's Republic of China}
 \affiliation{State Key Laboratory of Artificial Microstructure and Mesoscopic Physics and Frontiers Science Center for Nano-Optoelectronics, Peking University, Beijing 100871, People's Republic of China}

\date{\today}

\begin{abstract}
$\rm La_3Ni_2O_7$, a nickelate compound with a reported superconducting transition temperature of $\rm 80~K$, has attracted significant attention in recent years. %Despite intensive research, the microscopic mechanism underlying its superconductivity remains a topic of debate. 
Density-wave phenomena arising from strong electron correlations are widely regarded as key to unraveling the superconductivity mechanism, but the ordering and stability of these density waves remain a subject of contention in existing theoretical studies. In this work, we employ the variational Monte Carlo (VMC) method to thoroughly examine the nature of density waves as functions of Coulomb repulsion and exchange interactions in bilayer two-orbital model proposed for the high pressure phase of $\rm La_3Ni_2O_7$. We analyse the spin and charge correlation functions in a wide range of parameter space, and delineate a schematic phase diagram that separates different density-wave ground states. Our results provide useful insights into the understanding of electron correlations in $\rm La_3Ni_2O_7$, and highlight the potential of VMC to elucidate its superconducting mechanism.
\end{abstract} 

\maketitle

\section{Introduction}

The discovery of superconductivity in $\rm{La_3Ni_2O_7}$ with a maximum superconducting transition temperature ($T_c$) of $\rm{80~K}$ has established nickelates as another high-profile class of transition-metal-based high-temperature superconductors \cite{sun2023signatures, wang2024normal}. Under applied pressure ranging from ambient to $\rm{10\,GPa}$, $\rm{La_3Ni_2O_7}$ undergoes a structural phase transition from the $Amam$ to $Fmmm$ to $I4/mmm$ symmetry, forming a bilayered perovskite structure that hosts superconductivity below $T_c$ \cite{sun2023signatures, Li2025Identification}. This unique bilayer structure distinguishes $\rm{La_3Ni_2O_7}$ from single-layer cuprates. Density functional theory (DFT) calculations have predicted that the electronic bands near the Fermi energy are predominantly composed of $\rm{Ni}$-$d_{3z^2-r^2}$, $\rm{Ni}$-$d_{x^2-y^2}$, and $\rm{O}$-$p$ orbitals \cite{luo2023bilayer,yang2024orbital}. The presence of the $d_{3z^2-r^2}$ orbital gives rise to interesting $s^{\pm}$ pairing \cite{luo2024high,liu2023s,PhysRevB.108.L140505,zhang2024structural}, sparking widespread interest in unraveling the strongly correlated electronic structure of this material.

Extensive research has been conducted on the superconductivity of this system, including studies on the competition between $d$-wave and $s^\pm$ pairing symmetries, the evolution of superconductivity under varying pressure, and the detrimental effects of apical-oxygen deficiencies on superconducting properties \cite{luo2024high,PhysRevLett.131.236002,wang2024normal}. 
Meanwhile, density wave phenomena, including the spin density wave (SDW) and charge density wave (CDW), could play pivotal roles in investigations of superconductivity \cite{luo2023bilayer,PhysRevB.103.195150,wang2016coexistence}. 
Hence, studying density wave correlations is crucial for deciphering pairing mechanisms in these materials, as spin/charge fluctuations associated with density-wave transitions can mediate unconventional superconductivity \cite{PhysRevB.37.5182,PhysRevB.39.293}.
More broadly, the interplay between superconductivity and density waves represents a ubiquitous theme in correlated electron systems \cite{yu2021unusual}. 
In $\rm{La_3Ni_2O_7}$, experimental and computational studies have probed SDW behavior at ambient pressure, where the results have converged to a SDW state with wave vector of $(\frac{\pi}{2},\frac{\pi}{2})$ \cite{ZHAO20251239,PhysRevLett.132.256503,Khasanov_2025,PhysRevB.111.184401}.
The existence of CDW at ambient pressure is also examined with reported results differing quite significantly from one to another \cite{Khasanov_2025,liu2023evidence,ni2025spin}.
However, for the high pressure phase, due to the lack of experimental methods, the existence and the properties of both SDW and CDW are seldom discussed, only recent ultrafast optical spectroscopy experiments have revealed potential density wave signatures in high pressure phases, implicating their potential connection to superconductivity\cite{meng2024density}. 
Overall, despite rapidly growing advances in studies of $\rm{La_3Ni_2O_7}$, the precise nature of the density waves is far from being comprehensively understood, warranting further theoretical examination.

%Various computational methods have been employed to investigate the property of $\rm{La_3Ni_2O_7}$. These methods can be broadly categorized into two groups: first-principles methods, including density functional theory (DFT) \cite{PhysRevB.110.L140508}, GW approximation \cite{PhysRevLett.131.206501}, and dynamical mean field theory (DMFT) \cite{PhysRevB.111.125111}; and theoretical methods, such as multi-orbital random-phase-approximation (RPA) method \cite{liu2023s}, fluctuation-exchange (FLEX) approximation \cite{PhysRevLett.132.106002}, and functional renormalization group (FRG) \cite{PhysRevB.108.L140505}. However, both have limitations. First-principles methods can investigate materials based on their atomic structure without empirical parameters, but they often struggle to fully capture strongly correlated effects. Theoretical methods, on the other hand, are typically based on models derived from first-principles calculations and are widely used in strongly correlated electronic systems. However, they are constrained by their perturbative nature and arbitrary selection of parameters.

% 此为上一段的改动后版本，突出了扫U-J的意义和各个计算方法的用途.
Theoretical investigations of $\rm{La_3Ni_2O_7}$ can be broadly categorized into two approaches: first-principles methods and theoretical models. First-principles methods, including density functional theory (DFT) \cite{PhysRevB.110.L140508}, GW approximation \cite{PhysRevLett.131.206501}, and dynamical mean field theory (DMFT) \cite{PhysRevB.111.125111}, are widely used to derive effective model, calculate band structures, and analyze Fermi surfaces. 
First principles methods have a big advantage of investigating the pressure effects in a consistent framework by optimizing the structures at various pressures, but they also face significant challenges in fully capturing strong correlation effects.
Meanwhile, to establish an improved description of correlation effects, theoretical models are solved using various numerical methods, such as multi-orbital random-phase-approximation (RPA) method \cite{liu2023s}, fluctuation-exchange (FLEX) approximation \cite{PhysRevLett.132.106002}, and functional renormalization group (FRG) \cite{PhysRevB.108.L140505}.
%to analyze the $s^{\pm}$ pairing and superconductivity properties. 
These approaches, however, are not free of problems, especially in the field of $\rm{La_3Ni_2O_7}$ research: (i) One effective model is often not sufficient to cover different crystalline structures obtained at different pressures; (ii) models are often solved at a specific set of parameters aimed at explaining experimental observations, but different studies may employ different model forms and parameters, which are not trivially determined; (iii) the numerical methods employed to solve these models have accuracy and approximation limitations.
%constrained by the perturbative nature of the numerical methods employed or limited by arbitrary selection of parameters aimed at explaining experimental observations.
%When investigating the quantum phase transition that requires calculating the ground state across a wide range of parameters.  

In this work, we focus on examining the density wave correlations by applying the variational Monte Carlo (VMC) method that addresses the challenge of describing strong electron correlations \cite{misawa2019mvmc,tahara2008variational}.
Our theoretical foundation is the existing bilayer two-orbital model suitable for high pressure $\rm{La_3Ni_2O_7}$ \cite{luo2023bilayer}, but we vary the interaction parameters to obtain general insights into density wave correlations in this model.
%to study the bilayer two-orbital model proposed for high pressure $\rm{La_3Ni_2O_7}$\cite{luo2023bilayer}. 
VMC, with carefully constructed wavefunction ansatzes, has been successfully applied to study both prototypical models \cite{kato2020many} and real materials \cite{PhysRevB.99.245155}, and here a generalized pair wave function ansatz is devised to handle both the singlet and the triplet pairing, aimed at an unbiased treatment of correlation effects. 
We systematically investigate the possibility of SDW and CDW correlations and determine the possible wave vectors associated with each density wave. 
%Eventually, a comprehensive understanding of density waves in this model is established.
As a first step toward a deeper understanding of $\rm{La_3Ni_2O_7}$ phases, this work does not directly probe the 
superconductivity, but the behaviors of density waves can readily offer useful insights into the understanding of this system.

\section{Model and Method}

\begin{figure}
    \centering
    \includegraphics[width=\linewidth]{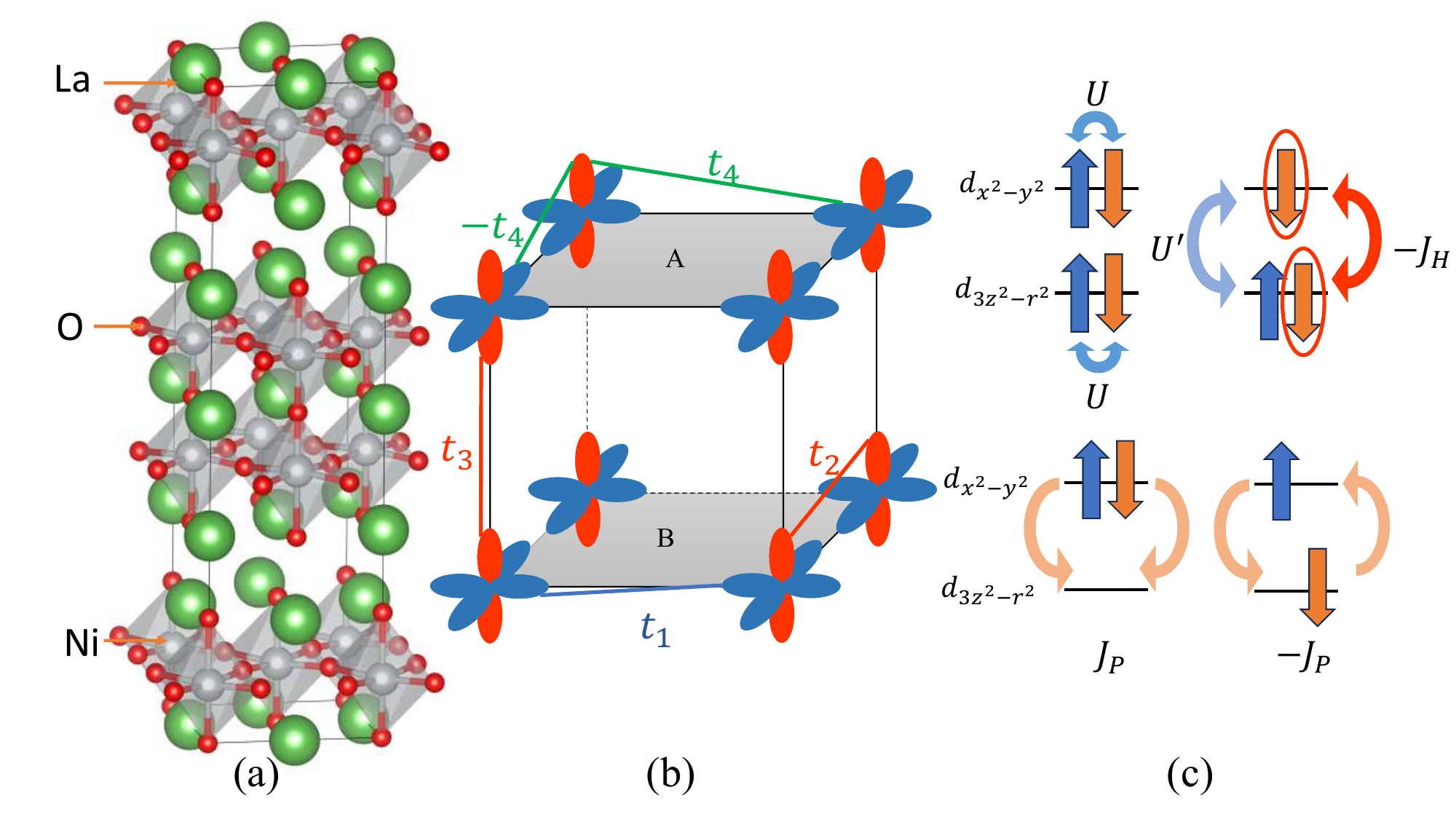}
    \caption{(a) The unit cell of $\rm La_3Ni_2O_7$, where $\rm La$, $\rm Ni$, and $\rm O$ atoms are represented by green, gray, and red spheres, respectively. (b) An illustration of the bilayer two-orbital tight-binding model. A and B label the two layers. The $\rm{Ni}$-$d_{3z^2-r^2}$ orbitals are depicted in red, while the $\rm{Ni}$-$d_{x^2-y^2}$ orbitals are shown in blue. The red, blue, and green lines represent the possible hopping terms in the model, with their corresponding values ($t_1$, $t_2$, $t_3$ and $t_4$) listed in Table \ref{tab1}. (c) A schematic illustration of all Coulomb interaction terms. The intra-orbital and inter-orbital Coulomb interactions are labeled as $U$ and $U^{\prime}$, respectively. The Hund's coupling between two electrons in the red-circled orbitals is denoted as $-J_H$, and the two types of pair-hopping interactions are labeled as $J_P$.}
    \label{fig1}
\end{figure}

\subsection{Model}

$\rm La_3Ni_2O_7$ has attracted considerable interest because of its unique bilayer structure hosted in its high-pressure phase, as shown in Fig.~\ref{fig1}(a). In contrast to cuprate high-temperature superconductors, which contain only a single $\rm Cu-O$ layer, $\rm La_3Ni_2O_7$ features two $\rm Ni-O$ layers connected by oxygen atoms. In this work, we employ the bilayer two-orbital ($d_{3z^2-r^2}$ and $d_{x^2-y^2}$) tight-binding (TB) model proposed for this high-pressure phase (Fig.~\ref{fig1}(b)) \cite{luo2023bilayer}. In this model, $\rm{Ni}$-$d_{3z^2-r^2}$ and $\rm{Ni}$-$d_{x^2-y^2}$ are placed on each site. The filling factor, defined as the number of electrons on each site, is 1.5. The Hamiltonian is shown as follows:
\begin{gather}
H_0 = - \sum_{\sigma}\sum_{ij} \sum_{\gamma} (t_1 c_{i \gamma 0 \sigma}^{\dagger} c_{j \gamma 0 \sigma} + t_2 c_{i \gamma 1 \sigma}^{\dagger} c_{j \gamma 1 \sigma} \nonumber\\+ t_4^{ij} c_{i \gamma 0 \sigma}^{\dagger} c_{j \gamma 1 \sigma} + \text{h.c.}) - \sum_{\sigma}\sum_i (t_3 c_{i A 1 \sigma}^{\dagger} c_{i B 1 \sigma} + \text{h.c.}) \\+ \sum_i \sum_\gamma \left(\epsilon_0 n_{i\gamma 0} + \epsilon_1 n_{i\gamma 1}\right) \nonumber
\end{gather}

Here, $i/j$ denotes different nearest-neighbor sites within the layer, $\sigma$ labels the spin, $\gamma$ labels the two distinct layers (A and B), and $0/1$ labels the $d_{x^2-y^2}$ orbital and $d_{3z^2-r^2}$ orbital, respectively. $c_{i\gamma\alpha\sigma}$ ($c_{i\gamma\alpha\sigma}^{\dagger}$) annihilates (creates) an electron with spin $\sigma$, located at orbital $\alpha$, site $i$ and layer $\gamma$. $n_{i\gamma\alpha}$ denotes the number of electrons located at orbital $\alpha$, site $i$ and layer $\gamma$. $t_1$, $t_2$, $t_3$, $t_4^{ij}$, $\epsilon_0$, $\epsilon_1$ are the hopping parameters and the on-site energy of orbitals. Specifically, the hopping parameter $t_4^{ij}$ can take values of $\pm t_4$ depending on the specific ${ij}$ pair, as illustrated in Fig. \ref{fig1}(b). All parameters ($t_1$, $t_2$, $t_3$, $t_4$, $\epsilon_0$, $\epsilon_1$) were obtained from the Wannier downfolding of the DFT band structure at $\rm{29\,GPa}$, as summarized in Table \ref{tab1}.

\begin{table}[b]%The best place to locate the table environment is directly after its first reference in text
\caption{\label{tab1}%
The hopping parameters between $d_{x^2-y^2}$ and $d_{3z^2-r^2}$ orbitals and their on-site energies. $t_1$, $t_2$, $t_3$ and $t_4$ are from Ref. \cite{PhysRevLett.132.106002}. $\epsilon_0$ and $\epsilon_1$ are from Ref. \cite{luo2023bilayer}. All data are in units of eV. As discussed in SI.I, this choice of parameters successfully rebuilds the DFT band structure.
}
\begin{ruledtabular}
\begin{tabular}{cccccc}
\textrm{$t_1$}&
\textrm{$t_2$}&
\textrm{$t_3$}&
\textrm{$t_4$}&
\textrm{$\epsilon_0$}&
\textrm{$\epsilon_1$}\\
\colrule
-0.491 & -0.117 & -0.664 & 0.242 & 0.776 & 0.409 \\
\end{tabular}
\end{ruledtabular}
\end{table}

We adopt the following Hamiltonian to describe the multi-orbital Coulomb interaction\cite{gu2023effective,PhysRevB.18.4945}:

\begin{gather}
H_1 = U \sum_i \sum_\gamma (n_{i \gamma 0 \uparrow} n_{i \gamma 0 \downarrow} + n_{i \gamma 1 \uparrow} n_{i \gamma 1 \downarrow})\nonumber\\
+U^{\prime} \sum_i \sum_{\gamma} n_{i \gamma 0} n_{i \gamma 1} - J_H \sum_i \sum_{\gamma} \sum_{\sigma} n_{i \gamma 0 \sigma} n_{i \gamma 1 \sigma} \nonumber\\ -J_P \sum_i \sum_{\gamma} \sum_\sigma (c_{i\gamma 0 \sigma}^\dagger c_{i\gamma 0 \bar{\sigma}}c_{i\gamma 1\bar{\sigma}}^\dagger c_{i\gamma 1\sigma}  \\  -c_{i\gamma 0\sigma}^\dagger c_{i\gamma 0\bar{\sigma}}^\dagger c_{i\gamma1\sigma}c_{i\gamma1\bar{\sigma}} + h.c.) \nonumber
\end{gather}
where $n_{i\gamma\alpha\sigma}$ denotes the number of electrons with spin $\sigma$, located at orbital $\alpha$, site $i$ and layer $\gamma$. 

In this Hamiltonian, the $U$($U^{\prime}$) term represents the intra(inter)-orbital Coulomb repulsion, the $J_H$ term denotes the Hund's coupling, and the $J_P$ term describes the pair-hopping interaction. The physical meaning of each term is illustrated in Fig. \ref{fig1}(c). Throughout this paper, we employ the Kanamori relations, which enforce $J_H = J_P = J$ and $U = U^{\prime} + 2J$\cite{PhysRevB.18.4945}.

To circumvent arbitrary parameter selection and investigate possible quantum phase transitions, in this study we vary values of $U = 3, 4, 5, 6, 8 \rm eV$ and $J/U = 0.05, 0.1, 0.15, 0.2$. These values span the parameter ranges frequently employed in relevant research and provide a sufficiently wide scope for our investigation.

\subsection{Variational Monte Carlo Calculation}

VMC is employed to solve the ground state of the model. VMC is based on the optimization of the wave function within the given ansatz. A more expressive ansatz has the potential to yield more accurate results, i.e. approaching closer to the true ground state. In the VMC calculation of a single-band Hubbard model, a commonly adopted wave function ansatz $|\Psi\rangle$ is constructed as the product of the following three terms: the Gutzwiller factor ($P_G$), the Jastrow factor ($P_J$), and the anti-parallel Pfaffian part ($|\psi_0\rangle$):
\begin{align}
|\Psi\rangle = P_G P_J |\psi_0\rangle
\end{align}
Each term is defined as follows:
\begin{gather}
        P_G = \exp\left(-\sum_{i}\frac{1}{2}g_{i} n_{i}(n_{i}-1)\right) \\
        P_J = \exp\left(\frac{1}{2}\sum_{i}\sum_{j}v_{ij}n_{i}n_{j}\right) \\
        |\psi_0\rangle = \left(\sum_{ij}f_{ij}c^{\dagger}_{i\uparrow}c^{\dagger}_{j\downarrow}\right)^{\frac{N_e}{2}} |0\rangle
\end{gather}
where $i$ and $j$ denote different sites, $ |0\rangle$ denotes the vacuum state, and $N_e$ represents the total number of electrons. 
Extensions of the ansatz to multi-band systems\cite{biborski2024variational,PhysRevB.84.180513}, systems with fixed total spin\cite{tahara2008variational}, and systems without the half-filling condition\cite{kato2020many}, have also been reported.

In the ansatz wavefunction proposed above, the Pfaffian part is generated by $\frac{N_e}{2}$ Cooper pairs which consist of two electrons with opposite spins, indicating a singlet superconductivity. Without losing generality, we employ a generalized pairing wave function as $|\psi_0\rangle$, which incorporates the bilayer two-orbital structure and possible triplet Cooper pairs\cite{tahara2008variational}:
\begin{gather}
        P_G = \exp\left(-\sum_{i\gamma\alpha}\frac{1}{2}g_{i\gamma\alpha} n_{i\gamma\alpha}(n_{i\gamma\alpha}-1)\right) \\
        P_J = \exp\left(\frac{1}{2}\sum_{i\gamma\alpha}\sum_{j\eta\beta}v^{i\gamma\alpha}_{j\eta\beta}n_{i\gamma\alpha}n_{j\eta\beta}\right) \\
        |\psi_0\rangle = \left(\sum_{\sigma\sigma^{\prime}}\sum_{i\gamma\alpha,j\eta\beta}F^{\sigma\sigma^{\prime}}_{i\gamma\alpha,j\eta\beta}c^{\dagger}_{i\gamma\alpha\sigma}c^{\dagger}_{j\eta\beta\sigma^{\prime}}\right)^{\frac{N_e}{2}} |0\rangle
\end{gather}
where $\alpha$ and $\beta$ label different orbitals. 

In our calculations, we utilize a square lattice with a side length ($N$) of 8, resulting in a total of $8 \times 8 \times 2$ sites and 192 electrons. Without loss of generality, $v^{i\gamma\alpha}_{j\eta\beta}$ and $F^{\sigma\sigma^{\prime}}_{i\gamma\alpha,j\eta\beta}$ are assumed to possess a $4 \times 4$ sublattice structure. The sublattice structure is defined as SM Eqs. S1 and S2.

To optimize the wave function in VMC, we employ the
minimum-step stochastic reconfiguration (MinSR) method \cite{rende2024simple}.
MinSR is modified against the standard stochastic reconfiguration (SR) method \cite{PhysRevB.64.024512}, where  
%several optimization methods have been developed, including stochastic gradient descent (SGD), conjugate gradient (CG), and the power Lanczos method\cite{heeb1993systematic,li2024convergenceanalysisstochasticgradient,PhysRevB.85.045103}. Among these, the most widely used approach is the stochastic reconfiguration (SR) method\cite{PhysRevB.64.024512}. In SR optimization, 
the overlap matrix $S$ and the gradient vector $\boldsymbol{g}$ are computed. These quantities are defined as follows:
\begin{equation}
\begin{aligned}
    S_{kl} = \langle O_k O_l \rangle - \langle O_k \rangle \langle O_l \rangle \\
    g_k = 2\langle O_k H \rangle - 2 \langle O_k \rangle \langle H \rangle
\end{aligned}
\end{equation}
where $O_k$ is the derivative operator, defined as $O_k(x) = \frac{1}{\langle x|\psi\rangle} \frac{\partial}{\partial \alpha_k}{\langle x|\psi\rangle}$, and $\alpha_k$ represents the $k$-th variational parameter.
Then the variational parameters can be updated by the following formula, with $\delta \tau$ controlling the step size of the optimization:
\begin{align}
    \boldsymbol{\Delta \alpha} = - \delta \tau S^{-1} \boldsymbol{g}
\end{align}

In our ansatz, the number of variational parameters ($N_p$) is large, and the $\mathcal{O}(N_p^3)$ computational complexity of the SR method becomes infeasible. 
In MinSR, a new 
%Recently, an improved SR method called the minimum-step SR (MinSR) has been proposed\cite{rende2024simple}. This method introduces a new 
matrix $Y$ and a new vector $\boldsymbol{f}$ are defined as:
\begin{equation}
\begin{aligned}
    Y_{kl} = O_{kl} - \langle O_k \rangle \\
    f_l = 2(H_l - \langle H\rangle)
\end{aligned}
\end{equation}
where $l$ denotes the $l$-th sample in sampling step. And the optimization can be carried out as follows:
\begin{equation}
\begin{aligned}
    \boldsymbol{\Delta \alpha} = -\delta \tau Y (Y^TY)^{-1}\boldsymbol{f}
\end{aligned}
\end{equation}

MinSR is equivalent to SR, but it can reduce the complexity to $\mathcal{O}(N_pN_s^2+N_s^3)$, where $N_s$ is the number of samples, which is usually much smaller than $N_p$. 
%It is usually more sufficient to optimize ansatz with large number of parameters, like neural-network ansatz\cite{chen2024empowering}. Here, we employ MinSR to optimize the ansatz wavefunction.

\section{Results}

In the study of density waves, one of the most critical aspects is determining the wave vector, as it may be related to the nesting of the Fermi surface in superconducting materials \cite{PhysRevB.77.165135,nakamura19951h,PhysRevB.110.L140508}. Recent computational results suggest the possibility of density waves with well-known wave vectors, such as $(\pi,\pi)$, $(\pi,0)$, and $(\frac{\pi}{2},\frac{\pi}{2})$\cite{luo2023bilayer,zhang2024structural,PhysRevMaterials.8.L111801} in $\rm{La_3Ni_2O_7}$ at different conditions and settings. 
In this work, we aim to investigate the possibility of density wave correlations in the studied model and identify their corresponding wave vectors.
Due to the bilayer structure of our model, the possible density wave configurations must account for inter-layer structure, which could be antiferromagnetic (AFM) or ferromagnetic (FM) order in the case of SDW. In previous studies of SDW, AFM order is often assumed due to the significant inter-layer hopping of the $d_{3z^2-r^2}$ orbital. However, the near half-filling condition of the $d_{3z^2-r^2}$ orbital also raises the possibility of FM order\cite{zhang2024structural}. 
Here, we will systematically examine the actual inter-layer structure to clarify this issue.

\subsection{Spin density wave}

\begin{figure*}
    \centering
    \includegraphics[width=\linewidth]{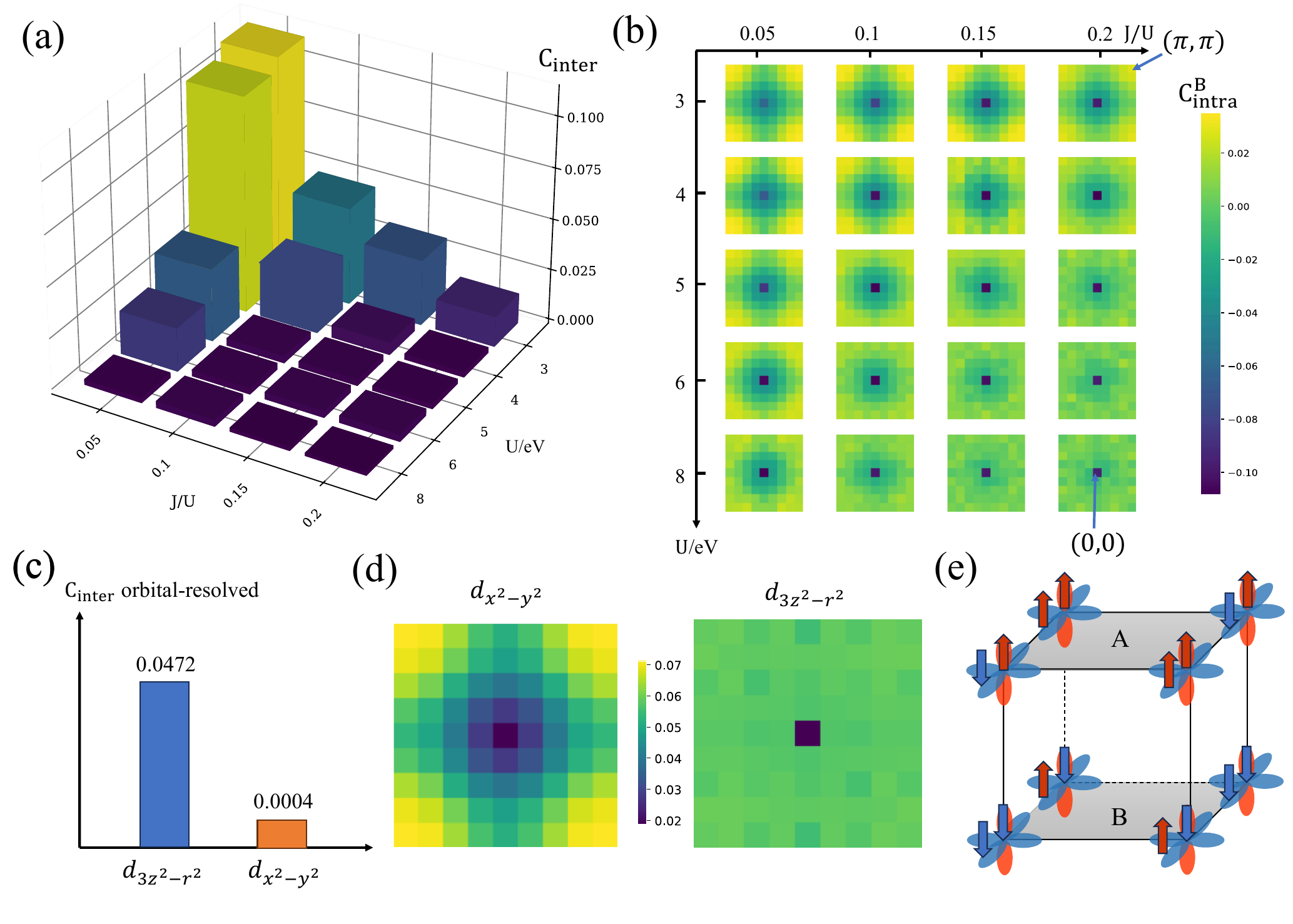}
    %\caption{Figure 2}
    \caption{Spin correlation analysis. (a) Values of $C_{\text{inter}}$ at different values of $U$ and $J/U$ in units of eV. (b) Values of $C_{\text{intra}}^{B}(\boldsymbol{q})$ for each $\boldsymbol{q}$. Each subfigure represents the $C_{\text{intra}}^{B}(\boldsymbol{q})$ distribution at a given $(U, J/U)$, with $q_x$ and $q_y$ ranging from $-\pi$ to $\pi$. The blue arrows highlight the points $\boldsymbol{q} = (\pi, \pi)$ and $\boldsymbol{q} = (0, 0)$. (c) Orbital-resolved inter-layer spin correlation function value ($C_{\text{inter}}^{\alpha}$) at $U=3$ and $J/U=0.1$. (d) Orbital-resolved intra-layer spin correlation function ($C_{\text{intra}}^{\prime\alpha\beta}(\boldsymbol{q})$) at $U=3$ and $J/U=0.1$. (e) Orbital-resolved schematic diagram of the spin configuration.}
    \label{fig2}
\end{figure*}

\begin{figure*}
    \centering
    \includegraphics[width=\linewidth]{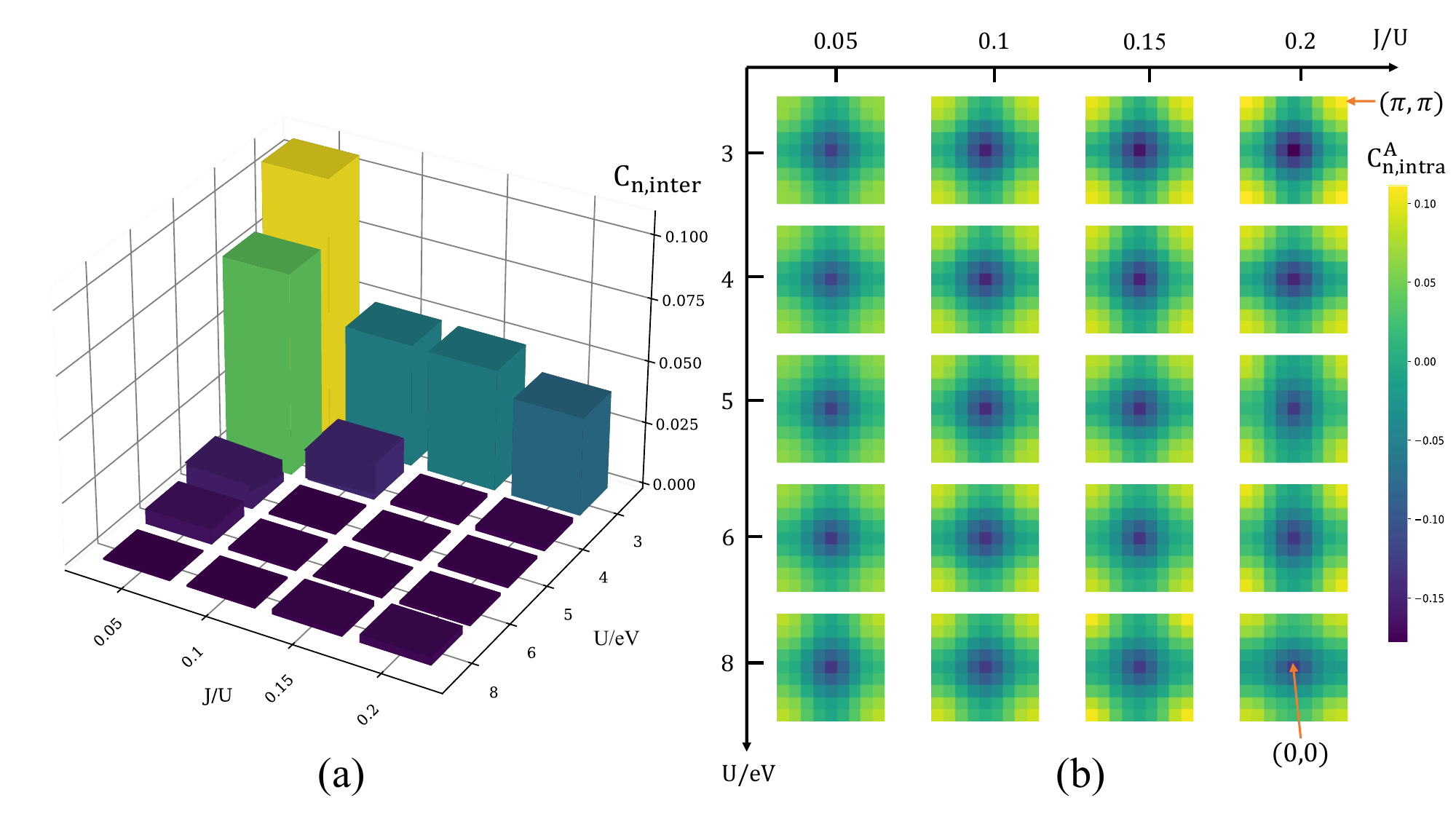}
    \caption{Correlation analysis of the charge density. Here, $U$ is in units of eV. (a) Values of $C_{n,\text{inter}}$ for each $(U, J/U)$ pair are shown. Note that $C_{n,\text{inter}}$ is positive when both $U$ and $J/U$ are small but decreases dramatically to nearly zero as $U$ and $J/U$ increase. (b) Values of $C_{n,\text{intra}}^{A}(\boldsymbol{q})$ for each $\boldsymbol{q}$ and $(U, J/U)$ pair. Each subfigure represents the $C_{n,\text{intra}}^{A}(\boldsymbol{q})$ distribution at a given $(U, J/U)$, with $q_x$ and $q_y$ ranging from $-\pi$ to $\pi$. The orange arrows highlight the points $\boldsymbol{q} = (\pi, \pi)$ and $\boldsymbol{q} = (0, 0)$.}
    \label{fig3}
\end{figure*}

To examine the inter-layer order for different values of $U$ and $J$, we employ the following real-space inter-layer correlation function, defined as:
\begin{align}
    C_{\text{inter}} = -\frac{1}{N^2}\sum_{i}S^z_{iA} S^z_{iB}
\end{align}
where $S^z_{iA}$ ($S^z_{iB}$) denotes the $z$-component of the spin on site $i$ in layer A (B). If the inter-layer correlation favors AFM order, $C_{\text{inter}}$ will be positive; for FM order, it will be negative. Additionally, the magnitude of $C_{\text{inter}}$ reflects the strength of the AFM (FM) correlation. The results are presented in Fig. \ref{fig2}(a).

Furthermore, we utilize the single-layer spin-correlation function in momentum space, denoted as $C_{\text{intra}}^\gamma(\boldsymbol{q})$, to investigate the possible intra-layer SDW types across a wide range of parameters. The function $C_{\text{intra}}^\gamma(\boldsymbol{q})$ is defined as:
\begin{align}
C_{\text{intra}}^\gamma(\boldsymbol{q}) = \frac{1}{N^2}\sum_{i\ne j}S^z_{i\gamma} S^z_{j\gamma} e^{-i \boldsymbol{q}\cdot (\boldsymbol{r_i}- \boldsymbol{r_j})}
\end{align}
where $\boldsymbol{q}=(q_x,q_y)$ represents the in-plane wave vector, and $\boldsymbol{r_i}$ and $\boldsymbol{r_j}$ denote the positions of sites $i$ and $j$ in layer $\gamma$. The restriction $i \ne j$ in the summation does not affect the relative intensity of each correlation. The results for layer $B$ are shown in Fig. \ref{fig2}(b), and the results for layer $A$ exhibit a similar pattern.

The calculations across a wide range of parameters consistently reveal the following results: $C_{\text{inter}}$ is always positive, indicating that the inter-layer order favors AFM. $C_{\text{inter}}$ exhibits a steep decrease as $U$ and $J/U$ increase. Additionally, the results for $C_{\text{intra}}^B$ show that the correlation function has a single peak at $(\pi,\pi)$, which weakens as $U$ and $J/U$ grow. No other significant peaks are observed in the intra-layer correlation function.

The correlation function analyses can provide deeper insights into the cooperative roles of $d_{x^2-y^2}$ and $d_{3z^2-r^2}$ orbitals. While previous orbital-resolved studies have predominantly emphasized the $d_{3z^2-r^2}$ orbital's contribution to $s^{\pm}$ superconductivity and strong inter-layer exchange coupling, we find that the $d_{x^2-y^2}$ orbital has equally important influence. Further considering this orbital's smaller effective mass, as demonstrated in recent studies \cite{luo2024high}, it may contribute significantly to the superfluid density in the superconducting phase.
To quantitatively assess the orbital-specific effects, we introduce 
the orbital-resolved inter-layer spin correlation function value as,
\begin{align}
    C_{\text{inter}}^{\alpha} = -\frac{1}{N^2}\sum_{i} \left(S^{\alpha}_{iz,A} - \langle S_z^{\alpha} \rangle\right) \left(S^{\alpha}_{iz,B} - \langle S_z^{\alpha} \rangle\right),
\end{align}
%where $S^{\alpha}_{iz,A}$ and $S^{\alpha}_{iz,B}$ denote the $z$-component of the spin on orbital $\alpha$ at site $i$ of layer A and B, respectively, with $\langle S_z^{\alpha} \rangle$ representing the orbital-averaged value. 
and the orbital-resolved intra-layer spin correlation function,
\begin{align}
    C_{\text{intra}}^{\prime\alpha\beta}(\boldsymbol{q}) = \frac{1}{N^2} \sum_{i,j} (S_{iz}^\alpha S_{jz}^\beta  - \langle S_z^{\alpha} \rangle \langle S_z^{\beta} \rangle ) e^{-i\boldsymbol{q} \cdot (\boldsymbol{r_i} - \boldsymbol{r_j})},
\end{align}
where $\alpha,\beta$ denote orbital indices ($d_{x^2-y^2}$ or $d_{3z^2-r^2}$). The $\langle S_z^{\alpha} \rangle$ denotes the average spin on $\alpha$ orbitals.
%$i$ and $j$ label sites, and $S_{jz}^\alpha$ represents spin $z$-component for orbital $\alpha$ at site $i$. %The result for $d_{x^2-y^2}$-$d_{x^2-y^2}$ intra-layer spin correlation function $C_{\text{intra}}^{\prime 00} (\boldsymbol{q})$ are shown in Fig.~\ref{fig4}(b)
%Furthermore, we define 
%The corresponding results $C_{\text{inter}}^{1}$ for $d_{3z^2-r^2}$ are presented in Fig.~\ref{fig4}(a).
The corresponding results are ploted in Fig.~\ref{fig2}c-d, revealing 
the following two key features. 
First, the $d_{3z^2-r^2}$ orbital governs the inter-layer spin correlations, displaying strong antiferromagnetic correlations that mirror the behavior of the total inter-layer correlation function.
Second, the $d_{x^2-y^2}$ orbital dominates the intra-layer spin correlations, exhibiting a pronounced $(\pi,\pi)$ peak that accounts for the majority of the total spin correlation strength. 
These orbital-resolved correlations are synthesized in the orbital-resolved schematic spin distribution diagram shown in Fig.~\ref{fig2}(d), which illustrates the distinct spatial patterns arising from each orbital's contribution.

\subsection{Charge density wave}

Similar to the analysis of SDW, we investigate the possible intra-layer and inter-layer structures of CDW. We employ the real-space inter-layer charge correlation function and the intra-layer charge-correlation function in momentum space, defined as:
\begin{gather}
    C_{n,\text{inter}} = -\frac{1}{N^2}\sum_{i} (n_{iA} - \langle n \rangle) (n_{iB} - \langle n \rangle) \\
    C_{n,\text{intra}}^\gamma(\boldsymbol{q}) = \frac{1}{N^2}\sum_{i \ne j} (n_{i\gamma} n_{j\gamma} - \langle n_\gamma \rangle^2) e^{-i \boldsymbol{q} \cdot (\boldsymbol{r_i} - \boldsymbol{r_j})}
\end{gather}
where $n_{i\gamma}$ denotes the number of electrons on site $i$ in layer $\gamma$, $\langle n \rangle$ represents the average number of electrons per site, and $\langle n_\gamma \rangle$ is the average number of electrons per site on each layer. The results are presented in Fig. \ref{fig3}.

The results for $C_{n,\text{inter}}$ show a different pattern compared to the spin case: $C_{n,\text{inter}}$ remains positive for all $(U, J/U)$ values and exhibits a dramatic decrease as $U$ and $J/U$ increase. But, for $C_{n,\text{intra}}^A$, we observe a consistent peak located at $(\pi,\pi)$ appears for all $(U,J/U)$. 
While a persistent $(\pi,\pi)$ peak in the intra-layer correlation function indicates the ubiquitous emergence of CDW regardless of $U$ or $J/U$ values, the abrupt suppression of $C_{n,\text{inter}}$ suggests the existence of two distinct CDW phases with different inter-layer configurations.

\section{discussion}

\begin{figure}
    \includegraphics[width=\linewidth]{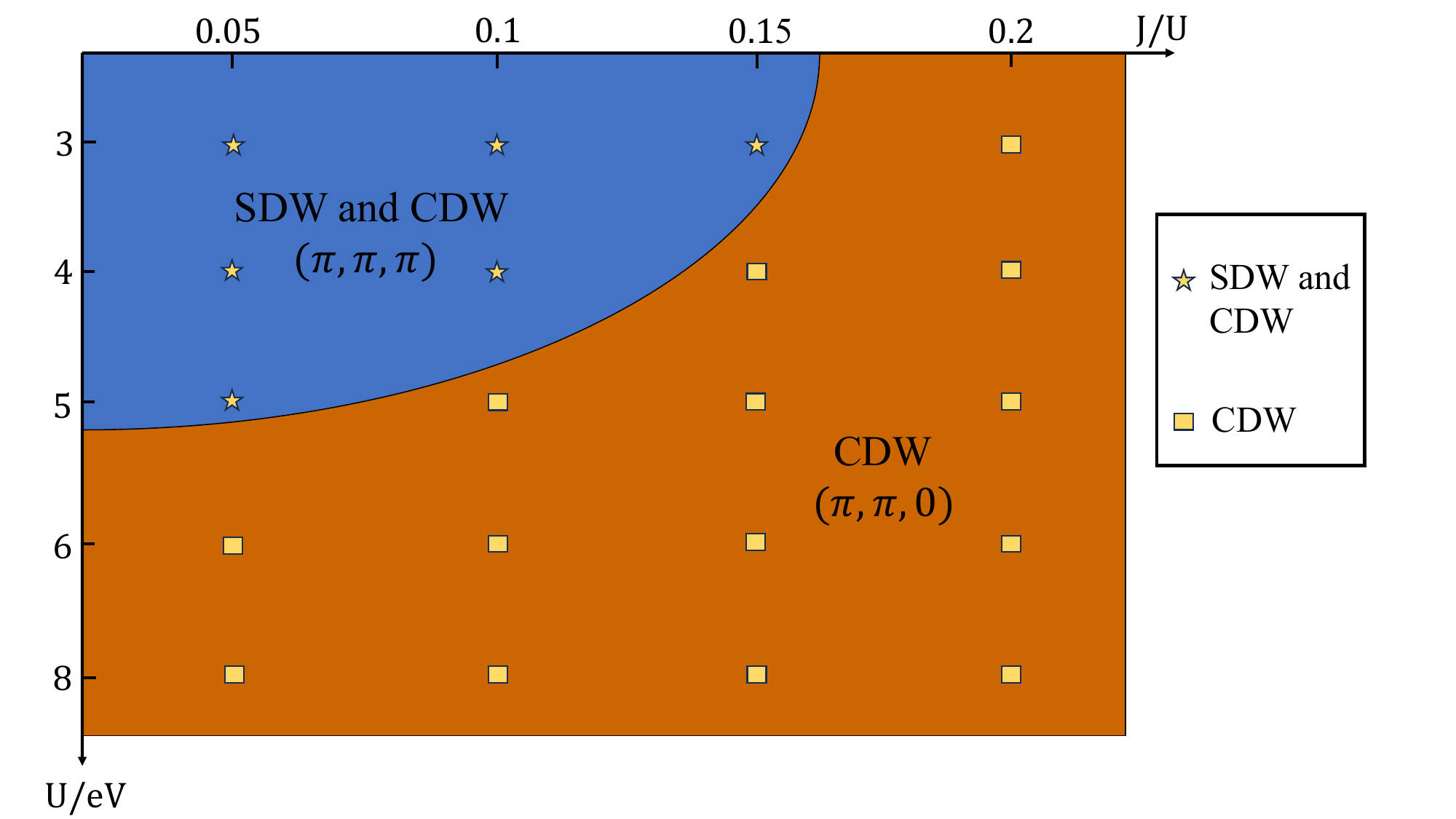}
    \caption{Schematic phase diagram illustrating the density-wave transition in $\rm{La_3Ni_2O_7}$ under high pressure. The phase diagram consists of two distinct regions: the region (blue) which SDW and CDW coexist and the CDW region (brown). For each $(U,J/U)$ pair examined,  we use yellow stars or squares to label their corresponding density-wave type. Within the coexisting region, both SDW and CDW orders are possible and may coexist, sharing the common wave vector $(\pi,\pi,\pi)$. As $U$ and $J/U$ increase, a phase transition occurs, leading to a CDW ground state with wave vector $(\pi,\pi,0)$}
    \label{fig4}
\end{figure}

From the investigation of the correlation functions, we can determine the wave vectors of SDW and CDW. For SDW, in the region of small $U$ and $J/U$, the positive $C_{\text{inter}}$ and the single peak at $(\pi,\pi)$ in the intra-layer correlation function indicte the SDW type in this region is the G-type antiferromagnetic (G-AFM), a state typically observed in bilayered SDW systems. 
This state is also conventionally denoted as the $(\pi,\pi,\pi)$ SDW.
As $U$ and $J/U$ increase, the $(\pi,\pi,\pi)$ SDW vanishes. 
Using the three-indices notation, for CDW, a similar analysis shows that the CDW wave vector is $(\pi,\pi,\pi)$ when $U$ and $J/U$ are both small. As $U$ and $J/U$ grow, the CDW wave vector undergoes a transition to $(\pi,\pi,0)$. These findings are summarized in the schematic phase diagram of SDW and CDW shown in Fig.~\ref{fig4}. 
Specifically, SDW appears only at relatively small $U$ and $J/U$, disappearing as these parameters increase. In contrast, CDW exists across all investigated parameter spaces, though its wave vector evolves with $U$ and $J/U$: $(\pi,\pi,\pi)$ at small values, shifting to $(\pi,\pi,0)$ at large values.
This suggests that SDW and CDW may coexist at small $U$ and $J/U$, while only CDW remains at large values.

The trends in density wave order can be rationalized by the following mechanisms: For SDW, the AFM exchange energy, scaling as $\sim t^2/U$, decreases with increasing $U$, while FM correlations driven by Hund's coupling $J$ become dominant\cite{PhysRevLett.127.077204}. 
This competition destabilizes of the SDW state as $U$ and $J/U$ increase. 
For intra-layer CDW, our findings resemble the physics revealed in extended Hubbard model, where CDW state is also favored as $J/U$ grows \cite{PhysRevLett.53.2327}. 
%For inter-layer CDW, enhanced Coulomb interaction ($U$) suppresses double occupations, while large $J/U$ promotes charge distribution across different orbitals.
%This, coupled with competition from intra-layer correlations, weakens inter-layer CDW.
The weakened inter-layer CDW behavior has been observed in recent DFT calculations, where this behavior is pressure-enhanced\cite{PhysRevB.110.L140508}. Their result might correspond to the relatively weakening $t_3$ compared to $t_1$, which indicates that the interplay between intra- and inter-layer CDW will play a role in the weakening inter-layer CDW behavior\cite{zhang2024structural}. Our result gives a good starting point for further investigation of it.

Our orbital analyses align with fundamental physics of multi-orbital systems. The $d_{x^2-y^2}$ orbital, characterized by a large in-plane hopping parameter $t_1$, generates substantial exchange energy $t_1^2/U$, dominating intra-layer spin correlations. 
This mechanism mirrors single-layer cuprates, where $d_{x^2-y^2}$ orbitals drive $(\pi,\pi)$ AFM correlations. 
Conversely, the $d_{3z^2-r^2}$ orbital exhibits distinct behavior through its large inter-layer hopping parameter $t_3$ arising from hybridization with O-$2p$ orbitals.
This enables strong inter-layer magnetic coupling, which was also revealed by DFT calculations \cite{sun2023signatures}. 
%This inter-orbital hybridization facilitates strong inter-layer magnetic coupling, explaining the pronounced $(\pi,\pi,\pi)$ correlations observed in our analysis. 
The electronic structure of $\rm La_3Ni_2O_7$ thus exhibits a dual character:
cuprate-like intra-layer physics governed by $d_{x^2-y^2}$ orbital correlations, and unique inter-layer AFM order mediated by $d_{3z^2-r^2}$ hybridization. 
These multi-band properties establish it as a versatile platform bridging features of both cuprates and iron-based superconductors.

Based on combined GW approximation and extended DMFT calculations, the high-pressure phase of $\rm{La_3Ni_2O_7}$ is estimated to have $U = 3–4$\,eV and $J = 0.61$\,eV \cite{PhysRevLett.131.206501}.
Within this parameter range, our results predict a density wave correlation that aligns with experimental observations\cite{meng2024density}.
The prediction is also consistent with previous DFT calculations, which attributed the observed density-wave-like behavior to CDW order\cite{PhysRevB.110.L140508}. 
Moreover, the predicted CDW-induced charge redistribution is consistent with the reported structural phase transition from DFT \cite{PhysRevMaterials.8.L111801}.
To directly probe the CDW order, future experimental studies could employ scanning tunneling microscopy, while structural characterization techniques like X-ray diffraction and electron microscopy can indirectly detect CDW-induced structural distortions \cite{PhysRevB.103.195126,CDW_STM,shi2024atomic}.
To this end, it is worth noting that, in realistic experiments, CDW-type orders may also arise from extrinsic factors such as impurities and sample inhomogeneities, hence care must be taken in experimental characterization and interpretation. 

It is worth noting that further algorithmic developments are needed to directly probe the superconductivity, including an accessible formula to compute superconductivity order within the employed VMC ansatz.
Nevertheless, the absence of SDW order in our calculations already provides some useful insights into the superconductivity of $\rm La_3Ni_2O_7$. Experimentally, SDW order is observed under ambient pressure, and it is suppressed as pressure increases. This suppression of SDW correlates with the onset of superconductivity, suggesting a spin-related mechanism underlying the superconducting state. 
Recently, there are also significant experimental and theoretical interests in deciphering the density wave behaviors at low-pressures. Although our VMC approach based on this wave function ansatz is suitable to shed light on this topic, a different model, originated from different crystalline structure and band structure, should be developed, and it is beyond the scope of this work.
%Furthermore, our calculations are based on a bilayer two-orbital tight-binding model, which is valid primarily for high-pressure structures. 
%To explore the density wave behavior at low pressures using our approach, a suitable effective model based on low pressure structures is desired.
%A more comprehensive investigation at ambient pressure requires the development of an effective model. 
%Furthermore, the presence of CDW order and the resulting structural phase transitions highlight the necessity of constructing a new theoretical model, which needs to be addressed in future work.
Moreover, the potential existence of incommensurate density-wave orders was predicted theoretically \cite{PhysRevLett.131.206501}, which also calls for further investigations in the future.

%Our analysis reveals three critical implications for understanding $\rm La_3Ni_2O_7$ under high pressure: First, the suppression of SDW order at GW+EDMFT derived interaction parameters mirrors the similar superconductivity origin in cuprates, where spin fluctuations are believed to mediate superconductivity. 
%Second, the predicted CDW-induced charge redistribution aligns with reported structural phase transitions from DFT \cite{PhysRevMaterials.8.L111801}. 
%However, such lattice distortions would fundamentally modify the model parameters—including hopping integrals $t_1$-$t_4$ and onsite energy $\epsilon_0$ and $\epsilon_1$, requiring a new theoretical model. 
%Third, while our lattice model identifies commensurate density-wave orders with wave vectors rational to the lattice periodicity, the potential existence of incommensurate density-wave phases—suggested by recent calculational results—highlights the need for deeper studies\cite{PhysRevB.110.L140508}.

\section{conclusion}

We employ VMC method to study the ground state of the bilayer two-orbital model of $\rm La_3N_2O_7$ under high pressure. 
A systematic exploration of a wide range of model parameters reveals that increasing $U$ and $J/U$ suppresses the $(\pi,\pi,\pi)$ SDW order while inducing a transition of the CDW wave vector from $(\pi,\pi,\pi)$ to $(\pi,\pi,0)$. 
The dominant $(\pi,\pi,0)$ CDW order identified in our calculations agrees well with both experimental observations and theoretical rationalizations, but the predicted SDW correlation vector in our high pressure phase-based model is different from the $(\pi/2, \pi/2)$ in-plane order observed in the low-pressure phase of $\rm La_3N_2O_7$.
This brings up an interesting question: how the SDW order change correlates with the structural phase transition upon applying external pressure?
To fully address this question, advanced \textit{ab initio} approaches that can properly treat electron correlations are necessary. 
This is beyond the scope of this work, but the present results suggest \textit{ab initio} VMC with a powerful ansatz could be a promising option.
In addition to the density wave orders, by analyzing orbital-resolved correlation functions, we elucidate the distinct roles of $d_{x^2-y^2}$ and $d_{3z^2-r^2}$ orbitals in shaping the system's electronic structure. 
This multi-orbital synergy gives rise to a unique orbital-resolved correlation effect, positioning $\rm La_3Ni_2O_7$ as a bridge between cuprates and iron-based superconducting materials.

\begin{acknowledgments}
The authors thank Tao Xiang, Daoxin Yao, and Kun Cao for helpful discussions.
This work was supported by the National Key R\&D Program of China under
Grant No. 2021YFA1400500, the Strategic Priority Research Program of the Chinese Academy of Sciences under Grant No. XDB33000000, and National Science Foundation of China under Grant No. 12334003. We are grateful for computational resources provided by the High Performance Computing Platform of Peking University.
\end{acknowledgments}

%\bibliographystyle{ieee}
%\printbibliography{}
\bibliography{ref}

\begin{thebibliography}{44}%
\makeatletter
\providecommand \@ifxundefined [1]{%
 \@ifx{#1\undefined}
}%
\providecommand \@ifnum [1]{%
 \ifnum #1\expandafter \@firstoftwo
 \else \expandafter \@secondoftwo
 \fi
}%
\providecommand \@ifx [1]{%
 \ifx #1\expandafter \@firstoftwo
 \else \expandafter \@secondoftwo
 \fi
}%
\providecommand \natexlab [1]{#1}%
\providecommand \enquote  [1]{``#1''}%
\providecommand \bibnamefont  [1]{#1}%
\providecommand \bibfnamefont [1]{#1}%
\providecommand \citenamefont [1]{#1}%
\providecommand \href@noop [0]{\@secondoftwo}%
\providecommand \href [0]{\begingroup \@sanitize@url \@href}%
\providecommand \@href[1]{\@@startlink{#1}\@@href}%
\providecommand \@@href[1]{\endgroup#1\@@endlink}%
\providecommand \@sanitize@url [0]{\catcode `\\12\catcode `\$12\catcode `\&12\catcode `\#12\catcode `\^12\catcode `\_12\catcode `\%12\relax}%
\providecommand \@@startlink[1]{}%
\providecommand \@@endlink[0]{}%
\providecommand \url  [0]{\begingroup\@sanitize@url \@url }%
\providecommand \@url [1]{\endgroup\@href {#1}{\urlprefix }}%
\providecommand \urlprefix  [0]{URL }%
\providecommand \Eprint [0]{\href }%
\providecommand \doibase [0]{http://dx.doi.org/}%
\providecommand \selectlanguage [0]{\@gobble}%
\providecommand \bibinfo  [0]{\@secondoftwo}%
\providecommand \bibfield  [0]{\@secondoftwo}%
\providecommand \translation [1]{[#1]}%
\providecommand \BibitemOpen [0]{}%
\providecommand \bibitemStop [0]{}%
\providecommand \bibitemNoStop [0]{.\EOS\space}%
\providecommand \EOS [0]{\spacefactor3000\relax}%
\providecommand \BibitemShut  [1]{\csname bibitem#1\endcsname}%
\let\auto@bib@innerbib\@empty
%</preamble>
\bibitem [{\citenamefont {Sun}\ \emph {et~al.}(2023)\citenamefont {Sun}, \citenamefont {Huo}, \citenamefont {Hu}, \citenamefont {Li}, \citenamefont {Liu}, \citenamefont {Han}, \citenamefont {Tang}, \citenamefont {Mao}, \citenamefont {Yang}, \citenamefont {Wang} \emph {et~al.}}]{sun2023signatures}%
  \BibitemOpen
  \bibfield  {author} {\bibinfo {author} {\bibfnamefont {H.}~\bibnamefont {Sun}}, \bibinfo {author} {\bibfnamefont {M.}~\bibnamefont {Huo}}, \bibinfo {author} {\bibfnamefont {X.}~\bibnamefont {Hu}}, \bibinfo {author} {\bibfnamefont {J.}~\bibnamefont {Li}}, \bibinfo {author} {\bibfnamefont {Z.}~\bibnamefont {Liu}}, \bibinfo {author} {\bibfnamefont {Y.}~\bibnamefont {Han}}, \bibinfo {author} {\bibfnamefont {L.}~\bibnamefont {Tang}}, \bibinfo {author} {\bibfnamefont {Z.}~\bibnamefont {Mao}}, \bibinfo {author} {\bibfnamefont {P.}~\bibnamefont {Yang}}, \bibinfo {author} {\bibfnamefont {B.}~\bibnamefont {Wang}},  \emph {et~al.},\ }\href@noop {} {\bibfield  {journal} {\bibinfo  {journal} {Nature}\ }\textbf {\bibinfo {volume} {621}},\ \bibinfo {pages} {493} (\bibinfo {year} {2023})}\BibitemShut {NoStop}%
\bibitem [{\citenamefont {Wang}\ \emph {et~al.}(2024)\citenamefont {Wang}, \citenamefont {Wen}, \citenamefont {Wu}, \citenamefont {Yao},\ and\ \citenamefont {Xiang}}]{wang2024normal}%
  \BibitemOpen
  \bibfield  {author} {\bibinfo {author} {\bibfnamefont {M.}~\bibnamefont {Wang}}, \bibinfo {author} {\bibfnamefont {H.-H.}\ \bibnamefont {Wen}}, \bibinfo {author} {\bibfnamefont {T.}~\bibnamefont {Wu}}, \bibinfo {author} {\bibfnamefont {D.-X.}\ \bibnamefont {Yao}}, \ and\ \bibinfo {author} {\bibfnamefont {T.}~\bibnamefont {Xiang}},\ }\href@noop {} {\bibfield  {journal} {\bibinfo  {journal} {Chinese Physics Letters}\ }\textbf {\bibinfo {volume} {41}},\ \bibinfo {pages} {077402} (\bibinfo {year} {2024})}\BibitemShut {NoStop}%
\bibitem [{\citenamefont {Li}\ \emph {et~al.}(2025)\citenamefont {Li}, \citenamefont {Peng}, \citenamefont {Ma}, \citenamefont {Zhang}, \citenamefont {Xing}, \citenamefont {Huang}, \citenamefont {Huang}, \citenamefont {Huo}, \citenamefont {Hu}, \citenamefont {Dong}, \citenamefont {Chen}, \citenamefont {Xie}, \citenamefont {Dong}, \citenamefont {Sun}, \citenamefont {Zeng}, \citenamefont {Mao},\ and\ \citenamefont {Wang}}]{Li2025Identification}%
  \BibitemOpen
  \bibfield  {author} {\bibinfo {author} {\bibfnamefont {J.}~\bibnamefont {Li}}, \bibinfo {author} {\bibfnamefont {D.}~\bibnamefont {Peng}}, \bibinfo {author} {\bibfnamefont {P.}~\bibnamefont {Ma}}, \bibinfo {author} {\bibfnamefont {H.}~\bibnamefont {Zhang}}, \bibinfo {author} {\bibfnamefont {Z.}~\bibnamefont {Xing}}, \bibinfo {author} {\bibfnamefont {X.}~\bibnamefont {Huang}}, \bibinfo {author} {\bibfnamefont {C.}~\bibnamefont {Huang}}, \bibinfo {author} {\bibfnamefont {M.}~\bibnamefont {Huo}}, \bibinfo {author} {\bibfnamefont {D.}~\bibnamefont {Hu}}, \bibinfo {author} {\bibfnamefont {Z.}~\bibnamefont {Dong}}, \bibinfo {author} {\bibfnamefont {X.}~\bibnamefont {Chen}}, \bibinfo {author} {\bibfnamefont {T.}~\bibnamefont {Xie}}, \bibinfo {author} {\bibfnamefont {H.}~\bibnamefont {Dong}}, \bibinfo {author} {\bibfnamefont {H.}~\bibnamefont {Sun}}, \bibinfo {author} {\bibfnamefont {Q.}~\bibnamefont {Zeng}}, \bibinfo {author} {\bibfnamefont {H.-k.}\ \bibnamefont {Mao}}, \ and\ \bibinfo {author} {\bibfnamefont
  {M.}~\bibnamefont {Wang}},\ }\href {\doibase 10.1093/nsr/nwaf220} {\bibfield  {journal} {\bibinfo  {journal} {National Science Review}\ ,\ \bibinfo {pages} {nwaf220}} (\bibinfo {year} {2025})}\BibitemShut {NoStop}%
\bibitem [{\citenamefont {Luo}\ \emph {et~al.}(2023)\citenamefont {Luo}, \citenamefont {Hu}, \citenamefont {Wang}, \citenamefont {W{\'u}},\ and\ \citenamefont {Yao}}]{luo2023bilayer}%
  \BibitemOpen
  \bibfield  {author} {\bibinfo {author} {\bibfnamefont {Z.}~\bibnamefont {Luo}}, \bibinfo {author} {\bibfnamefont {X.}~\bibnamefont {Hu}}, \bibinfo {author} {\bibfnamefont {M.}~\bibnamefont {Wang}}, \bibinfo {author} {\bibfnamefont {W.}~\bibnamefont {W{\'u}}}, \ and\ \bibinfo {author} {\bibfnamefont {D.-X.}\ \bibnamefont {Yao}},\ }\href@noop {} {\bibfield  {journal} {\bibinfo  {journal} {Physical review letters}\ }\textbf {\bibinfo {volume} {131}},\ \bibinfo {pages} {126001} (\bibinfo {year} {2023})}\BibitemShut {NoStop}%
\bibitem [{\citenamefont {Yang}\ \emph {et~al.}(2024)\citenamefont {Yang}, \citenamefont {Sun}, \citenamefont {Hu}, \citenamefont {Xie}, \citenamefont {Miao}, \citenamefont {Luo}, \citenamefont {Chen}, \citenamefont {Liang}, \citenamefont {Zhu}, \citenamefont {Qu} \emph {et~al.}}]{yang2024orbital}%
  \BibitemOpen
  \bibfield  {author} {\bibinfo {author} {\bibfnamefont {J.}~\bibnamefont {Yang}}, \bibinfo {author} {\bibfnamefont {H.}~\bibnamefont {Sun}}, \bibinfo {author} {\bibfnamefont {X.}~\bibnamefont {Hu}}, \bibinfo {author} {\bibfnamefont {Y.}~\bibnamefont {Xie}}, \bibinfo {author} {\bibfnamefont {T.}~\bibnamefont {Miao}}, \bibinfo {author} {\bibfnamefont {H.}~\bibnamefont {Luo}}, \bibinfo {author} {\bibfnamefont {H.}~\bibnamefont {Chen}}, \bibinfo {author} {\bibfnamefont {B.}~\bibnamefont {Liang}}, \bibinfo {author} {\bibfnamefont {W.}~\bibnamefont {Zhu}}, \bibinfo {author} {\bibfnamefont {G.}~\bibnamefont {Qu}},  \emph {et~al.},\ }\href@noop {} {\bibfield  {journal} {\bibinfo  {journal} {Nature Communications}\ }\textbf {\bibinfo {volume} {15}},\ \bibinfo {pages} {4373} (\bibinfo {year} {2024})}\BibitemShut {NoStop}%
\bibitem [{\citenamefont {Luo}\ \emph {et~al.}(2024)\citenamefont {Luo}, \citenamefont {Lv}, \citenamefont {Wang}, \citenamefont {W{\'u}},\ and\ \citenamefont {Yao}}]{luo2024high}%
  \BibitemOpen
  \bibfield  {author} {\bibinfo {author} {\bibfnamefont {Z.}~\bibnamefont {Luo}}, \bibinfo {author} {\bibfnamefont {B.}~\bibnamefont {Lv}}, \bibinfo {author} {\bibfnamefont {M.}~\bibnamefont {Wang}}, \bibinfo {author} {\bibfnamefont {W.}~\bibnamefont {W{\'u}}}, \ and\ \bibinfo {author} {\bibfnamefont {D.-X.}\ \bibnamefont {Yao}},\ }\href@noop {} {\bibfield  {journal} {\bibinfo  {journal} {npj Quantum Materials}\ }\textbf {\bibinfo {volume} {9}},\ \bibinfo {pages} {61} (\bibinfo {year} {2024})}\BibitemShut {NoStop}%
\bibitem [{\citenamefont {Liu}\ \emph {et~al.}(2023{\natexlab{a}})\citenamefont {Liu}, \citenamefont {Mei}, \citenamefont {Ye}, \citenamefont {Chen},\ and\ \citenamefont {Yang}}]{liu2023s}%
  \BibitemOpen
  \bibfield  {author} {\bibinfo {author} {\bibfnamefont {Y.-B.}\ \bibnamefont {Liu}}, \bibinfo {author} {\bibfnamefont {J.-W.}\ \bibnamefont {Mei}}, \bibinfo {author} {\bibfnamefont {F.}~\bibnamefont {Ye}}, \bibinfo {author} {\bibfnamefont {W.-Q.}\ \bibnamefont {Chen}}, \ and\ \bibinfo {author} {\bibfnamefont {F.}~\bibnamefont {Yang}},\ }\href@noop {} {\bibfield  {journal} {\bibinfo  {journal} {Physical Review Letters}\ }\textbf {\bibinfo {volume} {131}},\ \bibinfo {pages} {236002} (\bibinfo {year} {2023}{\natexlab{a}})}\BibitemShut {NoStop}%
\bibitem [{\citenamefont {Yang}\ \emph {et~al.}(2023)\citenamefont {Yang}, \citenamefont {Wang},\ and\ \citenamefont {Wang}}]{PhysRevB.108.L140505}%
  \BibitemOpen
  \bibfield  {author} {\bibinfo {author} {\bibfnamefont {Q.-G.}\ \bibnamefont {Yang}}, \bibinfo {author} {\bibfnamefont {D.}~\bibnamefont {Wang}}, \ and\ \bibinfo {author} {\bibfnamefont {Q.-H.}\ \bibnamefont {Wang}},\ }\href {\doibase 10.1103/PhysRevB.108.L140505} {\bibfield  {journal} {\bibinfo  {journal} {Phys. Rev. B}\ }\textbf {\bibinfo {volume} {108}},\ \bibinfo {pages} {L140505} (\bibinfo {year} {2023})}\BibitemShut {NoStop}%
\bibitem [{\citenamefont {Zhang}\ \emph {et~al.}(2024)\citenamefont {Zhang}, \citenamefont {Lin}, \citenamefont {Moreo}, \citenamefont {Maier},\ and\ \citenamefont {Dagotto}}]{zhang2024structural}%
  \BibitemOpen
  \bibfield  {author} {\bibinfo {author} {\bibfnamefont {Y.}~\bibnamefont {Zhang}}, \bibinfo {author} {\bibfnamefont {L.-F.}\ \bibnamefont {Lin}}, \bibinfo {author} {\bibfnamefont {A.}~\bibnamefont {Moreo}}, \bibinfo {author} {\bibfnamefont {T.~A.}\ \bibnamefont {Maier}}, \ and\ \bibinfo {author} {\bibfnamefont {E.}~\bibnamefont {Dagotto}},\ }\href@noop {} {\bibfield  {journal} {\bibinfo  {journal} {Nature Communications}\ }\textbf {\bibinfo {volume} {15}},\ \bibinfo {pages} {2470} (\bibinfo {year} {2024})}\BibitemShut {NoStop}%
\bibitem [{\citenamefont {Liu}\ \emph {et~al.}(2023{\natexlab{b}})\citenamefont {Liu}, \citenamefont {Mei}, \citenamefont {Ye}, \citenamefont {Chen},\ and\ \citenamefont {Yang}}]{PhysRevLett.131.236002}%
  \BibitemOpen
  \bibfield  {author} {\bibinfo {author} {\bibfnamefont {Y.-B.}\ \bibnamefont {Liu}}, \bibinfo {author} {\bibfnamefont {J.-W.}\ \bibnamefont {Mei}}, \bibinfo {author} {\bibfnamefont {F.}~\bibnamefont {Ye}}, \bibinfo {author} {\bibfnamefont {W.-Q.}\ \bibnamefont {Chen}}, \ and\ \bibinfo {author} {\bibfnamefont {F.}~\bibnamefont {Yang}},\ }\href {\doibase 10.1103/PhysRevLett.131.236002} {\bibfield  {journal} {\bibinfo  {journal} {Phys. Rev. Lett.}\ }\textbf {\bibinfo {volume} {131}},\ \bibinfo {pages} {236002} (\bibinfo {year} {2023}{\natexlab{b}})}\BibitemShut {NoStop}%
\bibitem [{\citenamefont {Lopes}\ \emph {et~al.}(2021)\citenamefont {Lopes}, \citenamefont {Reyes}, \citenamefont {Continentino},\ and\ \citenamefont {Thomas}}]{PhysRevB.103.195150}%
  \BibitemOpen
  \bibfield  {author} {\bibinfo {author} {\bibfnamefont {N.}~\bibnamefont {Lopes}}, \bibinfo {author} {\bibfnamefont {D.}~\bibnamefont {Reyes}}, \bibinfo {author} {\bibfnamefont {M.~A.}\ \bibnamefont {Continentino}}, \ and\ \bibinfo {author} {\bibfnamefont {C.}~\bibnamefont {Thomas}},\ }\href {\doibase 10.1103/PhysRevB.103.195150} {\bibfield  {journal} {\bibinfo  {journal} {Phys. Rev. B}\ }\textbf {\bibinfo {volume} {103}},\ \bibinfo {pages} {195150} (\bibinfo {year} {2021})}\BibitemShut {NoStop}%
\bibitem [{\citenamefont {Wang}\ \emph {et~al.}(2016)\citenamefont {Wang}, \citenamefont {Liu}, \citenamefont {Quan},\ and\ \citenamefont {Zou}}]{wang2016coexistence}%
  \BibitemOpen
  \bibfield  {author} {\bibinfo {author} {\bibfnamefont {Q.-W.}\ \bibnamefont {Wang}}, \bibinfo {author} {\bibfnamefont {D.-Y.}\ \bibnamefont {Liu}}, \bibinfo {author} {\bibfnamefont {Y.-M.}\ \bibnamefont {Quan}}, \ and\ \bibinfo {author} {\bibfnamefont {L.-J.}\ \bibnamefont {Zou}},\ }\href@noop {} {\bibfield  {journal} {\bibinfo  {journal} {Physics Letters A}\ }\textbf {\bibinfo {volume} {380}},\ \bibinfo {pages} {2685} (\bibinfo {year} {2016})}\BibitemShut {NoStop}%
\bibitem [{\citenamefont {Dong}\ \emph {et~al.}(1988)\citenamefont {Dong}, \citenamefont {Liang}, \citenamefont {Che}, \citenamefont {Xie}, \citenamefont {Zhao}, \citenamefont {Yang}, \citenamefont {Ni},\ and\ \citenamefont {Liu}}]{PhysRevB.37.5182}%
  \BibitemOpen
  \bibfield  {author} {\bibinfo {author} {\bibfnamefont {C.}~\bibnamefont {Dong}}, \bibinfo {author} {\bibfnamefont {J.~K.}\ \bibnamefont {Liang}}, \bibinfo {author} {\bibfnamefont {G.~C.}\ \bibnamefont {Che}}, \bibinfo {author} {\bibfnamefont {S.~S.}\ \bibnamefont {Xie}}, \bibinfo {author} {\bibfnamefont {Z.~X.}\ \bibnamefont {Zhao}}, \bibinfo {author} {\bibfnamefont {Q.~S.}\ \bibnamefont {Yang}}, \bibinfo {author} {\bibfnamefont {Y.~M.}\ \bibnamefont {Ni}}, \ and\ \bibinfo {author} {\bibfnamefont {G.~R.}\ \bibnamefont {Liu}},\ }\href {\doibase 10.1103/PhysRevB.37.5182} {\bibfield  {journal} {\bibinfo  {journal} {Phys. Rev. B}\ }\textbf {\bibinfo {volume} {37}},\ \bibinfo {pages} {5182} (\bibinfo {year} {1988})}\BibitemShut {NoStop}%
\bibitem [{\citenamefont {Tachiki}\ and\ \citenamefont {Takahashi}(1989)}]{PhysRevB.39.293}%
  \BibitemOpen
  \bibfield  {author} {\bibinfo {author} {\bibfnamefont {M.}~\bibnamefont {Tachiki}}\ and\ \bibinfo {author} {\bibfnamefont {S.}~\bibnamefont {Takahashi}},\ }\href {\doibase 10.1103/PhysRevB.39.293} {\bibfield  {journal} {\bibinfo  {journal} {Phys. Rev. B}\ }\textbf {\bibinfo {volume} {39}},\ \bibinfo {pages} {293} (\bibinfo {year} {1989})}\BibitemShut {NoStop}%
\bibitem [{\citenamefont {Yu}\ \emph {et~al.}(2021)\citenamefont {Yu}, \citenamefont {Ma}, \citenamefont {Zhuo}, \citenamefont {Liu}, \citenamefont {Wen}, \citenamefont {Lei}, \citenamefont {Ying},\ and\ \citenamefont {Chen}}]{yu2021unusual}%
  \BibitemOpen
  \bibfield  {author} {\bibinfo {author} {\bibfnamefont {F.}~\bibnamefont {Yu}}, \bibinfo {author} {\bibfnamefont {D.}~\bibnamefont {Ma}}, \bibinfo {author} {\bibfnamefont {W.}~\bibnamefont {Zhuo}}, \bibinfo {author} {\bibfnamefont {S.}~\bibnamefont {Liu}}, \bibinfo {author} {\bibfnamefont {X.}~\bibnamefont {Wen}}, \bibinfo {author} {\bibfnamefont {B.}~\bibnamefont {Lei}}, \bibinfo {author} {\bibfnamefont {J.}~\bibnamefont {Ying}}, \ and\ \bibinfo {author} {\bibfnamefont {X.}~\bibnamefont {Chen}},\ }\href@noop {} {\bibfield  {journal} {\bibinfo  {journal} {Nature communications}\ }\textbf {\bibinfo {volume} {12}},\ \bibinfo {pages} {3645} (\bibinfo {year} {2021})}\BibitemShut {NoStop}%
\bibitem [{\citenamefont {Zhao}\ \emph {et~al.}(2025)\citenamefont {Zhao}, \citenamefont {Zhou}, \citenamefont {Huo}, \citenamefont {Wang}, \citenamefont {Nie}, \citenamefont {Yang}, \citenamefont {Ying}, \citenamefont {Wang}, \citenamefont {Wu},\ and\ \citenamefont {Chen}}]{ZHAO20251239}%
  \BibitemOpen
  \bibfield  {author} {\bibinfo {author} {\bibfnamefont {D.}~\bibnamefont {Zhao}}, \bibinfo {author} {\bibfnamefont {Y.}~\bibnamefont {Zhou}}, \bibinfo {author} {\bibfnamefont {M.}~\bibnamefont {Huo}}, \bibinfo {author} {\bibfnamefont {Y.}~\bibnamefont {Wang}}, \bibinfo {author} {\bibfnamefont {L.}~\bibnamefont {Nie}}, \bibinfo {author} {\bibfnamefont {Y.}~\bibnamefont {Yang}}, \bibinfo {author} {\bibfnamefont {J.}~\bibnamefont {Ying}}, \bibinfo {author} {\bibfnamefont {M.}~\bibnamefont {Wang}}, \bibinfo {author} {\bibfnamefont {T.}~\bibnamefont {Wu}}, \ and\ \bibinfo {author} {\bibfnamefont {X.}~\bibnamefont {Chen}},\ }\href {\doibase https://doi.org/10.1016/j.scib.2025.02.019} {\bibfield  {journal} {\bibinfo  {journal} {Science Bulletin}\ }\textbf {\bibinfo {volume} {70}},\ \bibinfo {pages} {1239} (\bibinfo {year} {2025})}\BibitemShut {NoStop}%
\bibitem [{\citenamefont {Chen}\ \emph {et~al.}(2024)\citenamefont {Chen}, \citenamefont {Liu}, \citenamefont {Jiao}, \citenamefont {Zou}, \citenamefont {Jiang}, \citenamefont {Li}, \citenamefont {Luo}, \citenamefont {Wu}, \citenamefont {Zhang}, \citenamefont {Guo},\ and\ \citenamefont {Shu}}]{PhysRevLett.132.256503}%
  \BibitemOpen
  \bibfield  {author} {\bibinfo {author} {\bibfnamefont {K.}~\bibnamefont {Chen}}, \bibinfo {author} {\bibfnamefont {X.}~\bibnamefont {Liu}}, \bibinfo {author} {\bibfnamefont {J.}~\bibnamefont {Jiao}}, \bibinfo {author} {\bibfnamefont {M.}~\bibnamefont {Zou}}, \bibinfo {author} {\bibfnamefont {C.}~\bibnamefont {Jiang}}, \bibinfo {author} {\bibfnamefont {X.}~\bibnamefont {Li}}, \bibinfo {author} {\bibfnamefont {Y.}~\bibnamefont {Luo}}, \bibinfo {author} {\bibfnamefont {Q.}~\bibnamefont {Wu}}, \bibinfo {author} {\bibfnamefont {N.}~\bibnamefont {Zhang}}, \bibinfo {author} {\bibfnamefont {Y.}~\bibnamefont {Guo}}, \ and\ \bibinfo {author} {\bibfnamefont {L.}~\bibnamefont {Shu}},\ }\href {\doibase 10.1103/PhysRevLett.132.256503} {\bibfield  {journal} {\bibinfo  {journal} {Phys. Rev. Lett.}\ }\textbf {\bibinfo {volume} {132}},\ \bibinfo {pages} {256503} (\bibinfo {year} {2024})}\BibitemShut {NoStop}%
\bibitem [{\citenamefont {Khasanov}\ \emph {et~al.}(2025)\citenamefont {Khasanov}, \citenamefont {Hicken}, \citenamefont {Gawryluk}, \citenamefont {Sazgari}, \citenamefont {Plokhikh}, \citenamefont {Sorel}, \citenamefont {Bartkowiak}, \citenamefont {Bötzel}, \citenamefont {Lechermann}, \citenamefont {Eremin}, \citenamefont {Luetkens},\ and\ \citenamefont {Guguchia}}]{Khasanov_2025}%
  \BibitemOpen
  \bibfield  {author} {\bibinfo {author} {\bibfnamefont {R.}~\bibnamefont {Khasanov}}, \bibinfo {author} {\bibfnamefont {T.~J.}\ \bibnamefont {Hicken}}, \bibinfo {author} {\bibfnamefont {D.~J.}\ \bibnamefont {Gawryluk}}, \bibinfo {author} {\bibfnamefont {V.}~\bibnamefont {Sazgari}}, \bibinfo {author} {\bibfnamefont {I.}~\bibnamefont {Plokhikh}}, \bibinfo {author} {\bibfnamefont {L.~P.}\ \bibnamefont {Sorel}}, \bibinfo {author} {\bibfnamefont {M.}~\bibnamefont {Bartkowiak}}, \bibinfo {author} {\bibfnamefont {S.}~\bibnamefont {Bötzel}}, \bibinfo {author} {\bibfnamefont {F.}~\bibnamefont {Lechermann}}, \bibinfo {author} {\bibfnamefont {I.~M.}\ \bibnamefont {Eremin}}, \bibinfo {author} {\bibfnamefont {H.}~\bibnamefont {Luetkens}}, \ and\ \bibinfo {author} {\bibfnamefont {Z.}~\bibnamefont {Guguchia}},\ }\href {\doibase 10.1038/s41567-024-02754-z} {\bibfield  {journal} {\bibinfo  {journal} {Nature Physics}\ }\textbf {\bibinfo {volume} {21}},\ \bibinfo {pages} {430–436} (\bibinfo {year} {2025})}\BibitemShut
  {NoStop}%
\bibitem [{\citenamefont {Zhang}\ \emph {et~al.}(2025)\citenamefont {Zhang}, \citenamefont {Xu},\ and\ \citenamefont {Xiang}}]{PhysRevB.111.184401}%
  \BibitemOpen
  \bibfield  {author} {\bibinfo {author} {\bibfnamefont {B.}~\bibnamefont {Zhang}}, \bibinfo {author} {\bibfnamefont {C.}~\bibnamefont {Xu}}, \ and\ \bibinfo {author} {\bibfnamefont {H.}~\bibnamefont {Xiang}},\ }\href {\doibase 10.1103/PhysRevB.111.184401} {\bibfield  {journal} {\bibinfo  {journal} {Phys. Rev. B}\ }\textbf {\bibinfo {volume} {111}},\ \bibinfo {pages} {184401} (\bibinfo {year} {2025})}\BibitemShut {NoStop}%
\bibitem [{\citenamefont {Liu}\ \emph {et~al.}(2023{\natexlab{c}})\citenamefont {Liu}, \citenamefont {Sun}, \citenamefont {Huo}, \citenamefont {Ma}, \citenamefont {Ji}, \citenamefont {Yi}, \citenamefont {Li}, \citenamefont {Liu}, \citenamefont {Yu}, \citenamefont {Zhang} \emph {et~al.}}]{liu2023evidence}%
  \BibitemOpen
  \bibfield  {author} {\bibinfo {author} {\bibfnamefont {Z.}~\bibnamefont {Liu}}, \bibinfo {author} {\bibfnamefont {H.}~\bibnamefont {Sun}}, \bibinfo {author} {\bibfnamefont {M.}~\bibnamefont {Huo}}, \bibinfo {author} {\bibfnamefont {X.}~\bibnamefont {Ma}}, \bibinfo {author} {\bibfnamefont {Y.}~\bibnamefont {Ji}}, \bibinfo {author} {\bibfnamefont {E.}~\bibnamefont {Yi}}, \bibinfo {author} {\bibfnamefont {L.}~\bibnamefont {Li}}, \bibinfo {author} {\bibfnamefont {H.}~\bibnamefont {Liu}}, \bibinfo {author} {\bibfnamefont {J.}~\bibnamefont {Yu}}, \bibinfo {author} {\bibfnamefont {Z.}~\bibnamefont {Zhang}},  \emph {et~al.},\ }\href@noop {} {\bibfield  {journal} {\bibinfo  {journal} {Science China Physics, Mechanics \& Astronomy}\ }\textbf {\bibinfo {volume} {66}},\ \bibinfo {pages} {217411} (\bibinfo {year} {2023}{\natexlab{c}})}\BibitemShut {NoStop}%
\bibitem [{\citenamefont {Ni}\ \emph {et~al.}(2025)\citenamefont {Ni}, \citenamefont {Ji}, \citenamefont {He}, \citenamefont {Xie}, \citenamefont {Yao}, \citenamefont {Wang},\ and\ \citenamefont {Cao}}]{ni2025spin}%
  \BibitemOpen
  \bibfield  {author} {\bibinfo {author} {\bibfnamefont {X.-S.}\ \bibnamefont {Ni}}, \bibinfo {author} {\bibfnamefont {Y.}~\bibnamefont {Ji}}, \bibinfo {author} {\bibfnamefont {L.}~\bibnamefont {He}}, \bibinfo {author} {\bibfnamefont {T.}~\bibnamefont {Xie}}, \bibinfo {author} {\bibfnamefont {D.-X.}\ \bibnamefont {Yao}}, \bibinfo {author} {\bibfnamefont {M.}~\bibnamefont {Wang}}, \ and\ \bibinfo {author} {\bibfnamefont {K.}~\bibnamefont {Cao}},\ }\href@noop {} {\bibfield  {journal} {\bibinfo  {journal} {npj Quantum Materials}\ }\textbf {\bibinfo {volume} {10}},\ \bibinfo {pages} {17} (\bibinfo {year} {2025})}\BibitemShut {NoStop}%
\bibitem [{\citenamefont {Meng}\ \emph {et~al.}(2024)\citenamefont {Meng}, \citenamefont {Yang}, \citenamefont {Sun}, \citenamefont {Zhang}, \citenamefont {Luo}, \citenamefont {Chen}, \citenamefont {Ma}, \citenamefont {Wang}, \citenamefont {Hong}, \citenamefont {Wang} \emph {et~al.}}]{meng2024density}%
  \BibitemOpen
  \bibfield  {author} {\bibinfo {author} {\bibfnamefont {Y.}~\bibnamefont {Meng}}, \bibinfo {author} {\bibfnamefont {Y.}~\bibnamefont {Yang}}, \bibinfo {author} {\bibfnamefont {H.}~\bibnamefont {Sun}}, \bibinfo {author} {\bibfnamefont {S.}~\bibnamefont {Zhang}}, \bibinfo {author} {\bibfnamefont {J.}~\bibnamefont {Luo}}, \bibinfo {author} {\bibfnamefont {L.}~\bibnamefont {Chen}}, \bibinfo {author} {\bibfnamefont {X.}~\bibnamefont {Ma}}, \bibinfo {author} {\bibfnamefont {M.}~\bibnamefont {Wang}}, \bibinfo {author} {\bibfnamefont {F.}~\bibnamefont {Hong}}, \bibinfo {author} {\bibfnamefont {X.}~\bibnamefont {Wang}},  \emph {et~al.},\ }\href@noop {} {\bibfield  {journal} {\bibinfo  {journal} {Nature Communications}\ }\textbf {\bibinfo {volume} {15}},\ \bibinfo {pages} {10408} (\bibinfo {year} {2024})}\BibitemShut {NoStop}%
\bibitem [{\citenamefont {Yi}\ \emph {et~al.}(2024)\citenamefont {Yi}, \citenamefont {Meng}, \citenamefont {Li}, \citenamefont {Liao}, \citenamefont {Li}, \citenamefont {You}, \citenamefont {Gu},\ and\ \citenamefont {Su}}]{PhysRevB.110.L140508}%
  \BibitemOpen
  \bibfield  {author} {\bibinfo {author} {\bibfnamefont {X.-W.}\ \bibnamefont {Yi}}, \bibinfo {author} {\bibfnamefont {Y.}~\bibnamefont {Meng}}, \bibinfo {author} {\bibfnamefont {J.-W.}\ \bibnamefont {Li}}, \bibinfo {author} {\bibfnamefont {Z.-W.}\ \bibnamefont {Liao}}, \bibinfo {author} {\bibfnamefont {W.}~\bibnamefont {Li}}, \bibinfo {author} {\bibfnamefont {J.-Y.}\ \bibnamefont {You}}, \bibinfo {author} {\bibfnamefont {B.}~\bibnamefont {Gu}}, \ and\ \bibinfo {author} {\bibfnamefont {G.}~\bibnamefont {Su}},\ }\href {\doibase 10.1103/PhysRevB.110.L140508} {\bibfield  {journal} {\bibinfo  {journal} {Phys. Rev. B}\ }\textbf {\bibinfo {volume} {110}},\ \bibinfo {pages} {L140508} (\bibinfo {year} {2024})}\BibitemShut {NoStop}%
\bibitem [{\citenamefont {Christiansson}\ \emph {et~al.}(2023)\citenamefont {Christiansson}, \citenamefont {Petocchi},\ and\ \citenamefont {Werner}}]{PhysRevLett.131.206501}%
  \BibitemOpen
  \bibfield  {author} {\bibinfo {author} {\bibfnamefont {V.}~\bibnamefont {Christiansson}}, \bibinfo {author} {\bibfnamefont {F.}~\bibnamefont {Petocchi}}, \ and\ \bibinfo {author} {\bibfnamefont {P.}~\bibnamefont {Werner}},\ }\href {\doibase 10.1103/PhysRevLett.131.206501} {\bibfield  {journal} {\bibinfo  {journal} {Phys. Rev. Lett.}\ }\textbf {\bibinfo {volume} {131}},\ \bibinfo {pages} {206501} (\bibinfo {year} {2023})}\BibitemShut {NoStop}%
\bibitem [{\citenamefont {Ouyang}\ \emph {et~al.}(2025)\citenamefont {Ouyang}, \citenamefont {Wang}, \citenamefont {He},\ and\ \citenamefont {Lu}}]{PhysRevB.111.125111}%
  \BibitemOpen
  \bibfield  {author} {\bibinfo {author} {\bibfnamefont {Z.}~\bibnamefont {Ouyang}}, \bibinfo {author} {\bibfnamefont {J.-M.}\ \bibnamefont {Wang}}, \bibinfo {author} {\bibfnamefont {R.-Q.}\ \bibnamefont {He}}, \ and\ \bibinfo {author} {\bibfnamefont {Z.-Y.}\ \bibnamefont {Lu}},\ }\href {\doibase 10.1103/PhysRevB.111.125111} {\bibfield  {journal} {\bibinfo  {journal} {Phys. Rev. B}\ }\textbf {\bibinfo {volume} {111}},\ \bibinfo {pages} {125111} (\bibinfo {year} {2025})}\BibitemShut {NoStop}%
\bibitem [{\citenamefont {Sakakibara}\ \emph {et~al.}(2024)\citenamefont {Sakakibara}, \citenamefont {Kitamine}, \citenamefont {Ochi},\ and\ \citenamefont {Kuroki}}]{PhysRevLett.132.106002}%
  \BibitemOpen
  \bibfield  {author} {\bibinfo {author} {\bibfnamefont {H.}~\bibnamefont {Sakakibara}}, \bibinfo {author} {\bibfnamefont {N.}~\bibnamefont {Kitamine}}, \bibinfo {author} {\bibfnamefont {M.}~\bibnamefont {Ochi}}, \ and\ \bibinfo {author} {\bibfnamefont {K.}~\bibnamefont {Kuroki}},\ }\href {\doibase 10.1103/PhysRevLett.132.106002} {\bibfield  {journal} {\bibinfo  {journal} {Phys. Rev. Lett.}\ }\textbf {\bibinfo {volume} {132}},\ \bibinfo {pages} {106002} (\bibinfo {year} {2024})}\BibitemShut {NoStop}%
\bibitem [{\citenamefont {Misawa}\ \emph {et~al.}(2019)\citenamefont {Misawa}, \citenamefont {Morita}, \citenamefont {Yoshimi}, \citenamefont {Kawamura}, \citenamefont {Motoyama}, \citenamefont {Ido}, \citenamefont {Ohgoe}, \citenamefont {Imada},\ and\ \citenamefont {Kato}}]{misawa2019mvmc}%
  \BibitemOpen
  \bibfield  {author} {\bibinfo {author} {\bibfnamefont {T.}~\bibnamefont {Misawa}}, \bibinfo {author} {\bibfnamefont {S.}~\bibnamefont {Morita}}, \bibinfo {author} {\bibfnamefont {K.}~\bibnamefont {Yoshimi}}, \bibinfo {author} {\bibfnamefont {M.}~\bibnamefont {Kawamura}}, \bibinfo {author} {\bibfnamefont {Y.}~\bibnamefont {Motoyama}}, \bibinfo {author} {\bibfnamefont {K.}~\bibnamefont {Ido}}, \bibinfo {author} {\bibfnamefont {T.}~\bibnamefont {Ohgoe}}, \bibinfo {author} {\bibfnamefont {M.}~\bibnamefont {Imada}}, \ and\ \bibinfo {author} {\bibfnamefont {T.}~\bibnamefont {Kato}},\ }\href@noop {} {\bibfield  {journal} {\bibinfo  {journal} {Computer Physics Communications}\ }\textbf {\bibinfo {volume} {235}},\ \bibinfo {pages} {447} (\bibinfo {year} {2019})}\BibitemShut {NoStop}%
\bibitem [{\citenamefont {Tahara}\ and\ \citenamefont {Imada}(2008)}]{tahara2008variational}%
  \BibitemOpen
  \bibfield  {author} {\bibinfo {author} {\bibfnamefont {D.}~\bibnamefont {Tahara}}\ and\ \bibinfo {author} {\bibfnamefont {M.}~\bibnamefont {Imada}},\ }\href@noop {} {\bibfield  {journal} {\bibinfo  {journal} {Journal of the Physical Society of Japan}\ }\textbf {\bibinfo {volume} {77}},\ \bibinfo {pages} {114701} (\bibinfo {year} {2008})}\BibitemShut {NoStop}%
\bibitem [{\citenamefont {Kato}\ and\ \citenamefont {Kuroki}(2020)}]{kato2020many}%
  \BibitemOpen
  \bibfield  {author} {\bibinfo {author} {\bibfnamefont {D.}~\bibnamefont {Kato}}\ and\ \bibinfo {author} {\bibfnamefont {K.}~\bibnamefont {Kuroki}},\ }\href@noop {} {\bibfield  {journal} {\bibinfo  {journal} {Physical Review Research}\ }\textbf {\bibinfo {volume} {2}},\ \bibinfo {pages} {023156} (\bibinfo {year} {2020})}\BibitemShut {NoStop}%
\bibitem [{\citenamefont {Hirayama}\ \emph {et~al.}(2019)\citenamefont {Hirayama}, \citenamefont {Misawa}, \citenamefont {Ohgoe}, \citenamefont {Yamaji},\ and\ \citenamefont {Imada}}]{PhysRevB.99.245155}%
  \BibitemOpen
  \bibfield  {author} {\bibinfo {author} {\bibfnamefont {M.}~\bibnamefont {Hirayama}}, \bibinfo {author} {\bibfnamefont {T.}~\bibnamefont {Misawa}}, \bibinfo {author} {\bibfnamefont {T.}~\bibnamefont {Ohgoe}}, \bibinfo {author} {\bibfnamefont {Y.}~\bibnamefont {Yamaji}}, \ and\ \bibinfo {author} {\bibfnamefont {M.}~\bibnamefont {Imada}},\ }\href {\doibase 10.1103/PhysRevB.99.245155} {\bibfield  {journal} {\bibinfo  {journal} {Phys. Rev. B}\ }\textbf {\bibinfo {volume} {99}},\ \bibinfo {pages} {245155} (\bibinfo {year} {2019})}\BibitemShut {NoStop}%
\bibitem [{\citenamefont {Gu}\ \emph {et~al.}(2023)\citenamefont {Gu}, \citenamefont {Le}, \citenamefont {Yang}, \citenamefont {Wu},\ and\ \citenamefont {Hu}}]{gu2023effective}%
  \BibitemOpen
  \bibfield  {author} {\bibinfo {author} {\bibfnamefont {Y.}~\bibnamefont {Gu}}, \bibinfo {author} {\bibfnamefont {C.}~\bibnamefont {Le}}, \bibinfo {author} {\bibfnamefont {Z.}~\bibnamefont {Yang}}, \bibinfo {author} {\bibfnamefont {X.}~\bibnamefont {Wu}}, \ and\ \bibinfo {author} {\bibfnamefont {J.}~\bibnamefont {Hu}},\ }\href@noop {} {\bibfield  {journal} {\bibinfo  {journal} {arXiv preprint arXiv:2306.07275}\ } (\bibinfo {year} {2023})}\BibitemShut {NoStop}%
\bibitem [{\citenamefont {Castellani}\ \emph {et~al.}(1978)\citenamefont {Castellani}, \citenamefont {Natoli},\ and\ \citenamefont {Ranninger}}]{PhysRevB.18.4945}%
  \BibitemOpen
  \bibfield  {author} {\bibinfo {author} {\bibfnamefont {C.}~\bibnamefont {Castellani}}, \bibinfo {author} {\bibfnamefont {C.~R.}\ \bibnamefont {Natoli}}, \ and\ \bibinfo {author} {\bibfnamefont {J.}~\bibnamefont {Ranninger}},\ }\href {\doibase 10.1103/PhysRevB.18.4945} {\bibfield  {journal} {\bibinfo  {journal} {Phys. Rev. B}\ }\textbf {\bibinfo {volume} {18}},\ \bibinfo {pages} {4945} (\bibinfo {year} {1978})}\BibitemShut {NoStop}%
\bibitem [{\citenamefont {Biborski}\ \emph {et~al.}(2024)\citenamefont {Biborski}, \citenamefont {W{\'o}jcik},\ and\ \citenamefont {Zegrodnik}}]{biborski2024variational}%
  \BibitemOpen
  \bibfield  {author} {\bibinfo {author} {\bibfnamefont {A.}~\bibnamefont {Biborski}}, \bibinfo {author} {\bibfnamefont {P.}~\bibnamefont {W{\'o}jcik}}, \ and\ \bibinfo {author} {\bibfnamefont {M.}~\bibnamefont {Zegrodnik}},\ }\href@noop {} {\bibfield  {journal} {\bibinfo  {journal} {Physical Review B}\ }\textbf {\bibinfo {volume} {109}},\ \bibinfo {pages} {125144} (\bibinfo {year} {2024})}\BibitemShut {NoStop}%
\bibitem [{\citenamefont {Maier}\ and\ \citenamefont {Scalapino}(2011)}]{PhysRevB.84.180513}%
  \BibitemOpen
  \bibfield  {author} {\bibinfo {author} {\bibfnamefont {T.~A.}\ \bibnamefont {Maier}}\ and\ \bibinfo {author} {\bibfnamefont {D.~J.}\ \bibnamefont {Scalapino}},\ }\href {\doibase 10.1103/PhysRevB.84.180513} {\bibfield  {journal} {\bibinfo  {journal} {Phys. Rev. B}\ }\textbf {\bibinfo {volume} {84}},\ \bibinfo {pages} {180513} (\bibinfo {year} {2011})}\BibitemShut {NoStop}%
\bibitem [{\citenamefont {Rende}\ \emph {et~al.}(2024)\citenamefont {Rende}, \citenamefont {Viteritti}, \citenamefont {Bardone}, \citenamefont {Becca},\ and\ \citenamefont {Goldt}}]{rende2024simple}%
  \BibitemOpen
  \bibfield  {author} {\bibinfo {author} {\bibfnamefont {R.}~\bibnamefont {Rende}}, \bibinfo {author} {\bibfnamefont {L.~L.}\ \bibnamefont {Viteritti}}, \bibinfo {author} {\bibfnamefont {L.}~\bibnamefont {Bardone}}, \bibinfo {author} {\bibfnamefont {F.}~\bibnamefont {Becca}}, \ and\ \bibinfo {author} {\bibfnamefont {S.}~\bibnamefont {Goldt}},\ }\href@noop {} {\bibfield  {journal} {\bibinfo  {journal} {Communications Physics}\ }\textbf {\bibinfo {volume} {7}},\ \bibinfo {pages} {260} (\bibinfo {year} {2024})}\BibitemShut {NoStop}%
\bibitem [{\citenamefont {Sorella}(2001)}]{PhysRevB.64.024512}%
  \BibitemOpen
  \bibfield  {author} {\bibinfo {author} {\bibfnamefont {S.}~\bibnamefont {Sorella}},\ }\href {\doibase 10.1103/PhysRevB.64.024512} {\bibfield  {journal} {\bibinfo  {journal} {Phys. Rev. B}\ }\textbf {\bibinfo {volume} {64}},\ \bibinfo {pages} {024512} (\bibinfo {year} {2001})}\BibitemShut {NoStop}%
\bibitem [{\citenamefont {Johannes}\ and\ \citenamefont {Mazin}(2008)}]{PhysRevB.77.165135}%
  \BibitemOpen
  \bibfield  {author} {\bibinfo {author} {\bibfnamefont {M.~D.}\ \bibnamefont {Johannes}}\ and\ \bibinfo {author} {\bibfnamefont {I.~I.}\ \bibnamefont {Mazin}},\ }\href {\doibase 10.1103/PhysRevB.77.165135} {\bibfield  {journal} {\bibinfo  {journal} {Phys. Rev. B}\ }\textbf {\bibinfo {volume} {77}},\ \bibinfo {pages} {165135} (\bibinfo {year} {2008})}\BibitemShut {NoStop}%
\bibitem [{\citenamefont {Nakamura}\ \emph {et~al.}(1995)\citenamefont {Nakamura}, \citenamefont {Nobutoki}, \citenamefont {Kobayashi}, \citenamefont {Takahashi},\ and\ \citenamefont {Saito}}]{nakamura19951h}%
  \BibitemOpen
  \bibfield  {author} {\bibinfo {author} {\bibfnamefont {T.}~\bibnamefont {Nakamura}}, \bibinfo {author} {\bibfnamefont {T.}~\bibnamefont {Nobutoki}}, \bibinfo {author} {\bibfnamefont {Y.}~\bibnamefont {Kobayashi}}, \bibinfo {author} {\bibfnamefont {T.}~\bibnamefont {Takahashi}}, \ and\ \bibinfo {author} {\bibfnamefont {G.}~\bibnamefont {Saito}},\ }\href@noop {} {\bibfield  {journal} {\bibinfo  {journal} {Synthetic Metals}\ }\textbf {\bibinfo {volume} {70}},\ \bibinfo {pages} {1293} (\bibinfo {year} {1995})}\BibitemShut {NoStop}%
\bibitem [{\citenamefont {LaBollita}\ \emph {et~al.}(2024)\citenamefont {LaBollita}, \citenamefont {Pardo}, \citenamefont {Norman},\ and\ \citenamefont {Botana}}]{PhysRevMaterials.8.L111801}%
  \BibitemOpen
  \bibfield  {author} {\bibinfo {author} {\bibfnamefont {H.}~\bibnamefont {LaBollita}}, \bibinfo {author} {\bibfnamefont {V.}~\bibnamefont {Pardo}}, \bibinfo {author} {\bibfnamefont {M.~R.}\ \bibnamefont {Norman}}, \ and\ \bibinfo {author} {\bibfnamefont {A.~S.}\ \bibnamefont {Botana}},\ }\href {\doibase 10.1103/PhysRevMaterials.8.L111801} {\bibfield  {journal} {\bibinfo  {journal} {Phys. Rev. Mater.}\ }\textbf {\bibinfo {volume} {8}},\ \bibinfo {pages} {L111801} (\bibinfo {year} {2024})}\BibitemShut {NoStop}%
\bibitem [{\citenamefont {Lin}\ \emph {et~al.}(2021)\citenamefont {Lin}, \citenamefont {Zhang}, \citenamefont {Alvarez}, \citenamefont {Moreo},\ and\ \citenamefont {Dagotto}}]{PhysRevLett.127.077204}%
  \BibitemOpen
  \bibfield  {author} {\bibinfo {author} {\bibfnamefont {L.-F.}\ \bibnamefont {Lin}}, \bibinfo {author} {\bibfnamefont {Y.}~\bibnamefont {Zhang}}, \bibinfo {author} {\bibfnamefont {G.}~\bibnamefont {Alvarez}}, \bibinfo {author} {\bibfnamefont {A.}~\bibnamefont {Moreo}}, \ and\ \bibinfo {author} {\bibfnamefont {E.}~\bibnamefont {Dagotto}},\ }\href {\doibase 10.1103/PhysRevLett.127.077204} {\bibfield  {journal} {\bibinfo  {journal} {Phys. Rev. Lett.}\ }\textbf {\bibinfo {volume} {127}},\ \bibinfo {pages} {077204} (\bibinfo {year} {2021})}\BibitemShut {NoStop}%
\bibitem [{\citenamefont {Hirsch}(1984)}]{PhysRevLett.53.2327}%
  \BibitemOpen
  \bibfield  {author} {\bibinfo {author} {\bibfnamefont {J.~E.}\ \bibnamefont {Hirsch}},\ }\href {\doibase 10.1103/PhysRevLett.53.2327} {\bibfield  {journal} {\bibinfo  {journal} {Phys. Rev. Lett.}\ }\textbf {\bibinfo {volume} {53}},\ \bibinfo {pages} {2327} (\bibinfo {year} {1984})}\BibitemShut {NoStop}%
\bibitem [{\citenamefont {Lingannan}\ \emph {et~al.}(2021)\citenamefont {Lingannan}, \citenamefont {Joseph}, \citenamefont {Vajeeston}, \citenamefont {Kuo}, \citenamefont {Lue}, \citenamefont {Kalaiselvan}, \citenamefont {Rajak},\ and\ \citenamefont {Arumugam}}]{PhysRevB.103.195126}%
  \BibitemOpen
  \bibfield  {author} {\bibinfo {author} {\bibfnamefont {G.}~\bibnamefont {Lingannan}}, \bibinfo {author} {\bibfnamefont {B.}~\bibnamefont {Joseph}}, \bibinfo {author} {\bibfnamefont {P.}~\bibnamefont {Vajeeston}}, \bibinfo {author} {\bibfnamefont {C.~N.}\ \bibnamefont {Kuo}}, \bibinfo {author} {\bibfnamefont {C.~S.}\ \bibnamefont {Lue}}, \bibinfo {author} {\bibfnamefont {G.}~\bibnamefont {Kalaiselvan}}, \bibinfo {author} {\bibfnamefont {P.}~\bibnamefont {Rajak}}, \ and\ \bibinfo {author} {\bibfnamefont {S.}~\bibnamefont {Arumugam}},\ }\href {\doibase 10.1103/PhysRevB.103.195126} {\bibfield  {journal} {\bibinfo  {journal} {Phys. Rev. B}\ }\textbf {\bibinfo {volume} {103}},\ \bibinfo {pages} {195126} (\bibinfo {year} {2021})}\BibitemShut {NoStop}%
\bibitem [{\citenamefont {Shen}\ \emph {et~al.}(2025)\citenamefont {Shen}, \citenamefont {Xie}, \citenamefont {Li}, \citenamefont {Deng}, \citenamefont {Ma}, \citenamefont {Chen}, \citenamefont {Fu}, \citenamefont {Li}, \citenamefont {Yuan}, \citenamefont {Ji}, \citenamefont {He}, \citenamefont {Guan},\ and\ \citenamefont {Kong}}]{CDW_STM}%
  \BibitemOpen
  \bibfield  {author} {\bibinfo {author} {\bibfnamefont {J.}~\bibnamefont {Shen}}, \bibinfo {author} {\bibfnamefont {X.}~\bibnamefont {Xie}}, \bibinfo {author} {\bibfnamefont {W.}~\bibnamefont {Li}}, \bibinfo {author} {\bibfnamefont {C.}~\bibnamefont {Deng}}, \bibinfo {author} {\bibfnamefont {Y.}~\bibnamefont {Ma}}, \bibinfo {author} {\bibfnamefont {H.}~\bibnamefont {Chen}}, \bibinfo {author} {\bibfnamefont {H.}~\bibnamefont {Fu}}, \bibinfo {author} {\bibfnamefont {F.-S.}\ \bibnamefont {Li}}, \bibinfo {author} {\bibfnamefont {B.}~\bibnamefont {Yuan}}, \bibinfo {author} {\bibfnamefont {C.}~\bibnamefont {Ji}}, \bibinfo {author} {\bibfnamefont {R.}~\bibnamefont {He}}, \bibinfo {author} {\bibfnamefont {J.}~\bibnamefont {Guan}}, \ and\ \bibinfo {author} {\bibfnamefont {W.}~\bibnamefont {Kong}},\ }\href {\doibase 10.1126/sciadv.adr9753} {\bibfield  {journal} {\bibinfo  {journal} {Science Advances}\ }\textbf {\bibinfo {volume} {11}},\ \bibinfo {pages} {eadr9753} (\bibinfo {year} {2025})}\BibitemShut {NoStop}%
\bibitem [{\citenamefont {Shi}\ \emph {et~al.}(2024)\citenamefont {Shi}, \citenamefont {Li}, \citenamefont {Xu}, \citenamefont {Han}, \citenamefont {Zhu}, \citenamefont {Liu}, \citenamefont {Qi}, \citenamefont {Zhang}, \citenamefont {Du}, \citenamefont {Chen} \emph {et~al.}}]{shi2024atomic}%
  \BibitemOpen
  \bibfield  {author} {\bibinfo {author} {\bibfnamefont {R.}~\bibnamefont {Shi}}, \bibinfo {author} {\bibfnamefont {Q.}~\bibnamefont {Li}}, \bibinfo {author} {\bibfnamefont {X.}~\bibnamefont {Xu}}, \bibinfo {author} {\bibfnamefont {B.}~\bibnamefont {Han}}, \bibinfo {author} {\bibfnamefont {R.}~\bibnamefont {Zhu}}, \bibinfo {author} {\bibfnamefont {F.}~\bibnamefont {Liu}}, \bibinfo {author} {\bibfnamefont {R.}~\bibnamefont {Qi}}, \bibinfo {author} {\bibfnamefont {X.}~\bibnamefont {Zhang}}, \bibinfo {author} {\bibfnamefont {J.}~\bibnamefont {Du}}, \bibinfo {author} {\bibfnamefont {J.}~\bibnamefont {Chen}},  \emph {et~al.},\ }\href@noop {} {\bibfield  {journal} {\bibinfo  {journal} {Nature Communications}\ }\textbf {\bibinfo {volume} {15}},\ \bibinfo {pages} {3418} (\bibinfo {year} {2024})}\BibitemShut {NoStop}%
\end{thebibliography}%


\begin{thebibliography}{2}%
\makeatletter
\providecommand \@ifxundefined [1]{%
 \@ifx{#1\undefined}
}%
\providecommand \@ifnum [1]{%
 \ifnum #1\expandafter \@firstoftwo
 \else \expandafter \@secondoftwo
 \fi
}%
\providecommand \@ifx [1]{%
 \ifx #1\expandafter \@firstoftwo
 \else \expandafter \@secondoftwo
 \fi
}%
\providecommand \natexlab [1]{#1}%
\providecommand \enquote  [1]{``#1''}%
\providecommand \bibnamefont  [1]{#1}%
\providecommand \bibfnamefont [1]{#1}%
\providecommand \citenamefont [1]{#1}%
\providecommand \href@noop [0]{\@secondoftwo}%
\providecommand \href [0]{\begingroup \@sanitize@url \@href}%
\providecommand \@href[1]{\@@startlink{#1}\@@href}%
\providecommand \@@href[1]{\endgroup#1\@@endlink}%
\providecommand \@sanitize@url [0]{\catcode `\\12\catcode `\$12\catcode `\&12\catcode `\#12\catcode `\^12\catcode `\_12\catcode `\%12\relax}%
\providecommand \@@startlink[1]{}%
\providecommand \@@endlink[0]{}%
\providecommand \url  [0]{\begingroup\@sanitize@url \@url }%
\providecommand \@url [1]{\endgroup\@href {#1}{\urlprefix }}%
\providecommand \urlprefix  [0]{URL }%
\providecommand \Eprint [0]{\href }%
\providecommand \doibase [0]{http://dx.doi.org/}%
\providecommand \selectlanguage [0]{\@gobble}%
\providecommand \bibinfo  [0]{\@secondoftwo}%
\providecommand \bibfield  [0]{\@secondoftwo}%
\providecommand \translation [1]{[#1]}%
\providecommand \BibitemOpen [0]{}%
\providecommand \bibitemStop [0]{}%
\providecommand \bibitemNoStop [0]{.\EOS\space}%
\providecommand \EOS [0]{\spacefactor3000\relax}%
\providecommand \BibitemShut  [1]{\csname bibitem#1\endcsname}%
\let\auto@bib@innerbib\@empty
%</preamble>
\bibitem [{\citenamefont {Sakakibara}\ \emph {et~al.}(2024)\citenamefont {Sakakibara}, \citenamefont {Kitamine}, \citenamefont {Ochi},\ and\ \citenamefont {Kuroki}}]{PhysRevLett.132.106002}%
  \BibitemOpen
  \bibfield  {author} {\bibinfo {author} {\bibfnamefont {H.}~\bibnamefont {Sakakibara}}, \bibinfo {author} {\bibfnamefont {N.}~\bibnamefont {Kitamine}}, \bibinfo {author} {\bibfnamefont {M.}~\bibnamefont {Ochi}}, \ and\ \bibinfo {author} {\bibfnamefont {K.}~\bibnamefont {Kuroki}},\ }\href {\doibase 10.1103/PhysRevLett.132.106002} {\bibfield  {journal} {\bibinfo  {journal} {Phys. Rev. Lett.}\ }\textbf {\bibinfo {volume} {132}},\ \bibinfo {pages} {106002} (\bibinfo {year} {2024})}\BibitemShut {NoStop}%
\bibitem [{\citenamefont {Luo}\ \emph {et~al.}(2023)\citenamefont {Luo}, \citenamefont {Hu}, \citenamefont {Wang}, \citenamefont {W{\'u}},\ and\ \citenamefont {Yao}}]{luo2023bilayer}%
  \BibitemOpen
  \bibfield  {author} {\bibinfo {author} {\bibfnamefont {Z.}~\bibnamefont {Luo}}, \bibinfo {author} {\bibfnamefont {X.}~\bibnamefont {Hu}}, \bibinfo {author} {\bibfnamefont {M.}~\bibnamefont {Wang}}, \bibinfo {author} {\bibfnamefont {W.}~\bibnamefont {W{\'u}}}, \ and\ \bibinfo {author} {\bibfnamefont {D.-X.}\ \bibnamefont {Yao}},\ }\href@noop {} {\bibfield  {journal} {\bibinfo  {journal} {Physical review letters}\ }\textbf {\bibinfo {volume} {131}},\ \bibinfo {pages} {126001} (\bibinfo {year} {2023})}\BibitemShut {NoStop}%
\end{thebibliography}%

\end{document}

\title{Supplementary Material for Electron correlation driven density waves in high pressure $\rm{La_3Ni_2O_7}$}

\author{Yanxin Chen}
 \affiliation{School of Physics, Peking University, Beijing 100871, People's Republic of China.}
\author{Haoxiang Chen}
 \email{hxchen@pku.edu.cn}
 \affiliation{School of Physics, Peking University, Beijing 100871, People's Republic of China.}
\author{Tonghuan Jiang}
 \affiliation{School of Physics, Peking University, Beijing 100871, People's Republic of China.}
\author{Ji Chen}
 \email{ji.chen@pku.edu.cn}
 \affiliation{School of Physics, Peking University, Beijing 100871, People's Republic of China.}
 \affiliation{Interdisciplinary Institute of Light-Element Quantum Materials, Frontiers Science Center for Nano-Optoelectronics, Peking University, Beijing 100871, People's Republic of China}
 \affiliation{State Key Laboratory of Artificial Microstructure and Mesoscopic Physics and Frontiers Science Center for Nano-Optoelectronics, Peking University, Beijing 100871, People's Republic of China}

\date{\today}
\maketitle

\section{The band structure of the tight-binding model}

\begin{figure}[h]
    \centering
    \includegraphics[width=0.5\linewidth]{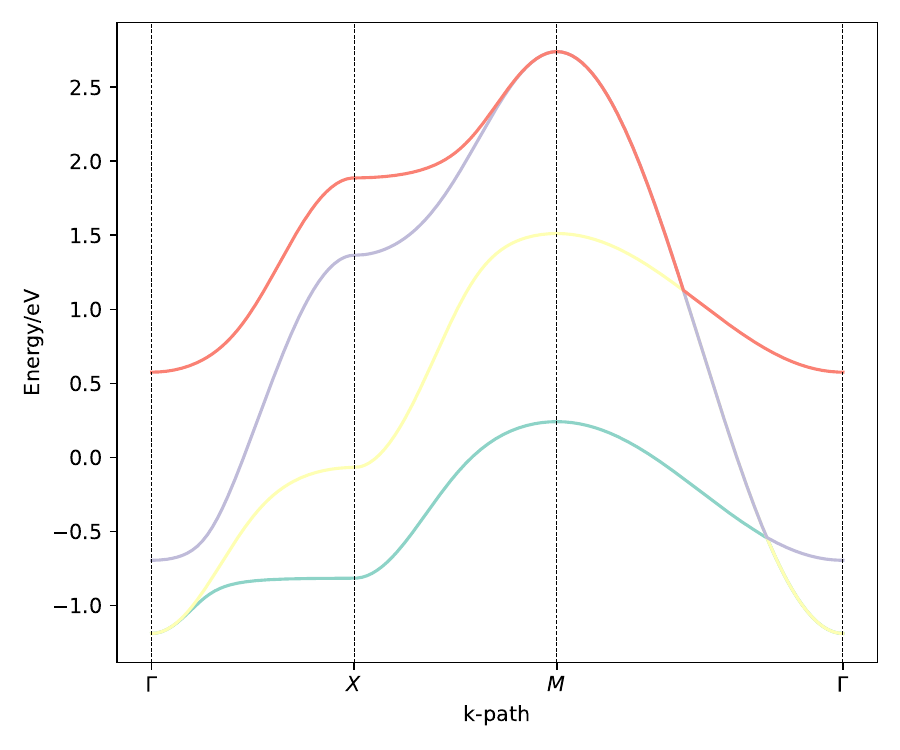}
    \caption{The band structure of the tight-binding model we employed. Each colored solid line represents a band. The model parameters are chosen to reproduce the DFT band structure. In Ref.\cite{PhysRevLett.132.106002}, a reasonable model was proposed, but the on-site energies $\epsilon_{0/1}$ were not given explicitly, and only their difference, $\Delta E = \epsilon_0 - \epsilon_1 = 0.372\mathrm{eV}$, was provided. Therefore, we adopt the on-site energies from Ref.\cite{luo2023bilayer}, which yield a very similar value, $\Delta E = 0.367\mathrm{eV}$. The reproduced band structure confirms the effectiveness of our model.}
    \label{SFig0}
\end{figure}

The band structure of the tight-binding model we employed is shown in Fig. \ref{SFig0}. It successfully reproduces the DFT band structure.

\section{The sublattice structure of ansatz parameters}

In our variational Monte Carlo (VMC) calculations, we adopt a lattice of side length $N = 8$. To reduce the number of variational parameters, we introduce an $N_s \times N_s$ sublattice structure (with $N_s = 4$) for the variational parameters $v^{i\gamma\alpha}_{j\eta\beta}$ and $F^{\sigma\sigma^{\prime}}_{i\gamma\alpha,j\eta\beta}$. These parameters are defined to depend on the relative vector $\boldsymbol{r_i} - \boldsymbol{r_j}$ and a sublattice index $g(i)$. Specifically, we express them as:
\begin{gather}
    v^{i\gamma\alpha}_{j\eta\beta} = v^{\gamma\eta,\alpha\beta}_{g(i)}(\boldsymbol{r_i} - \boldsymbol{r_j}), \\
    F^{\sigma\sigma^{\prime}}_{i\gamma\alpha,j\eta\beta} = F^{\sigma\sigma^{\prime},\gamma\eta,\alpha\beta}_{g(i)}(\boldsymbol{r_i} - \boldsymbol{r_j}).
\end{gather}

The sublattice index $g(i)$ for site $i$ is determined by:
\begin{gather}
    x = \left\lfloor \frac{i}{N} \right\rfloor, \quad y = i \mod N, \\
    g(i) = (x \mod N_s) \cdot N_s + (y \mod N_s).
\end{gather}

\section{Stabilization of SR and MinSR Methods}

In the optimization of the stochastic reconfiguration (SR) and minimum-step stochastic reconfiguration (MinSR) methods, the update of the variational parameters ($\Delta \boldsymbol{\alpha}$) follows Eqs.~\ref{Eqs1} and \ref{Eqs2}:
\begin{align}
    \Delta \boldsymbol{\alpha} = -\delta \tau S^{-1} \boldsymbol{g}, \label{Eqs1} \\
    \boldsymbol{\Delta \alpha} = -\delta \tau Y (Y^T Y)^{-1} \boldsymbol{f}, \label{Eqs2}
\end{align}
where $\boldsymbol{g}$ and $\boldsymbol{f}$ are gradient vectors, and $S$ and $Y$ are matrices. Here, $\delta \tau$ controls the step size of the optimization.

In both SR and MinSR optimizations, the inversion of matrices $S$ and $Y^T Y$ is required. A common issue is that these matrices may become ill-conditioned. To address this, we employ the following stabilization strategies:
\begin{itemize}
    \item Modification of the diagonal elements,
    \item Use of the pseudo-inverse with a cutoff for small singular values.
\end{itemize}

Specifically, when calculating the inversion of a matrix $A$, we stabilize the process by modifying $A$ to $A + \epsilon I$, where $\epsilon$ is a small positive real number. We then replace the inversion of $A + \epsilon I$ with its pseudo-inverse, computed using singular-value decomposition (SVD). The SVD procedure is as follows:
\begin{align}
    A + \epsilon I = U D V^\dagger,
\end{align}
where $U$ and $V$ are unitary matrices, and $D$ is a diagonal matrix whose elements are the singular values of $A + \epsilon I$, denoted as:
\begin{align}
    D = \text{diag}\{\lambda_1, \lambda_2, \lambda_3, \cdots, \lambda_{n-2}, \lambda_{n-1}, \lambda_n\}.
\end{align}
Since $A$ is a positive-definite matrix, all singular values are positive. Without loss of generality, we assume $\lambda_1 \ge \lambda_2 \ge \lambda_3 \ge \cdots \ge \lambda_n$. To stabilize the inversion, we truncate small singular values of $A + \epsilon I$ using the criterion that any singular value less than $\epsilon_{wf} \lambda_1$ is set to zero. After that, only $\lambda_1,\lambda_2,\cdots,\lambda_t$ remain. We denote the truncated inversion matrix as $\Tilde{D} = \rm diag\{\lambda_1^{-1}, \lambda_2^{-1},\cdots, \lambda_{t}^{-1},0,\cdots, 0\}$. The stabilized inverse of $A$ is then given by:
\begin{align}
    \bar{A}^{-1} = V \Tilde{D} U^\dagger.
\end{align}

In practice, we replace the theoretical formulas for SR and MinSR with the following stabilized versions:
\begin{align}
    \Delta \boldsymbol{\alpha} = -\delta \tau \bar{S}^{-1} \boldsymbol{g}, \label{Eqs3} \\
    \boldsymbol{\Delta \alpha} = -\delta \tau Y \bar{K}^{-1} \boldsymbol{f}, \label{Eqs4}
\end{align}
where $K = Y^T Y$.

\section{Strategies for generating uncorrelated samples}

In variational Monte Carlo calculations, we begin by proposing an ansatz wavefunction $|\Psi\rangle$. We then employ the Metropolis-Hastings algorithm to generate a set of statistically independent samples $\{\boldsymbol{x}\}$ distributed according to the probability density $\Psi(\boldsymbol{x}) = |\langle \boldsymbol{x}|\Psi\rangle|^2/\langle\Psi|\Psi\rangle$. 

The standard Metropolis-Hastings sampling procedure consists of the following steps:
\begin{enumerate}
    \item Start from a randomly chosen initial configuration $\boldsymbol{x_0}$
    \item Propose a new configuration $\boldsymbol{x^\prime}$ through a transition probability
    \item Accept the new configuration with probability $\min\left(1, \frac{|\langle \boldsymbol{x^\prime}|\Psi\rangle|^2}{|\langle \boldsymbol{x_0}|\Psi\rangle|^2}\right)$
\end{enumerate}

This sequence constitutes one Monte Carlo step. However, consecutive samples generated through this process exhibit strong correlations, which would lead to significant statistical errors if used directly in measurements.

To generate effectively uncorrelated samples, we implement two key strategies:
\begin{itemize}
    \item Discard the first $N_h$ samples to ensure the Markov chain reaches equilibrium
    \item Retain only every $N_i$-th sample to reduce autocorrelations
\end{itemize}

Through these procedures, we obtain $N_w$ statistically independent samples suitable for accurate measurements.

\section{The parameters employed in sampling and optimization}

The parameters applied in the optimization and sampling steps are shown in Tab. \ref{tab1} 

\begin{table}%The best place to locate the table environment is directly after its first reference in text
\caption{\label{tab1} A table shows the parameters applied in the optimization and sampling step, whose definitions are shown in Supplementary Information. $\delta t$, $\epsilon$ as well as $\epsilon_{wf}$ control the optimization, and $N_w$, $N_h$ also $N_i$ control the sampling step.  
}
\begin{ruledtabular}
\begin{tabular}{cccccc}
\textrm{$\delta \tau$}&
\textrm{$\epsilon$}&
\textrm{$\epsilon_{wf}$}&
\textrm{$N_w$}&
\textrm{$N_h$}&
\textrm{$N_i$}\\
\colrule
0.03 & 0.005 & 0.0005 & 3000 & 1500 & 192\\

\end{tabular}
\end{ruledtabular}
\end{table}

\begin{figure}
    \centering
    \includegraphics[width=\linewidth]{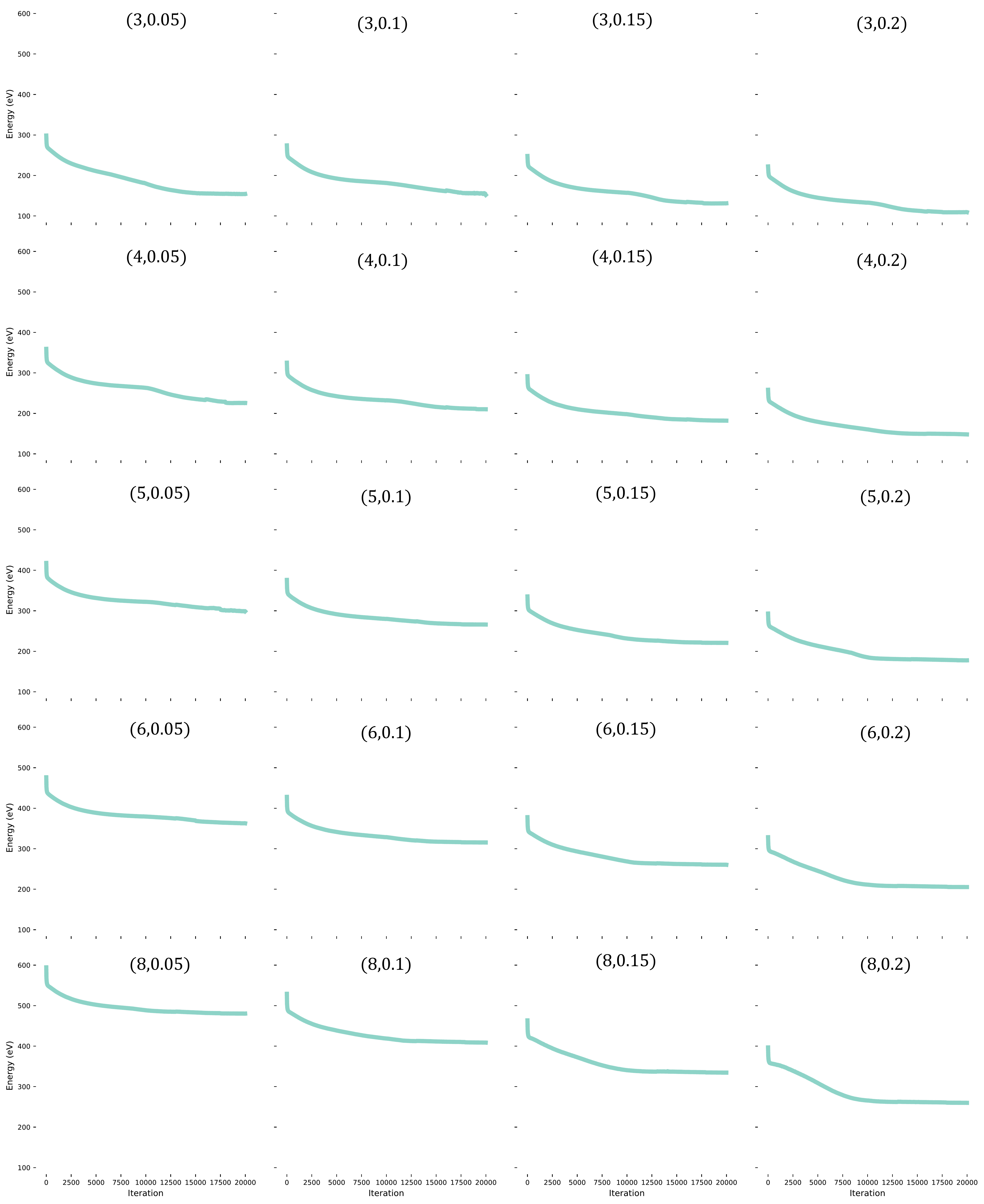}
    \caption{The Hamiltonian-iteration figure for every $(U,J/U)$ pairs. All 20 wave functions are converged.}
    \label{SFig2}
\end{figure}

After 20000 iterations, the optimization has converged. The Hamiltonian-iteration diagrams are shown in Fig. \ref{SFig2} 

\section{Intra-layer correlation function without the $i \ne j$ constraint}

The conventional definition of the single-layer correlation function is given by:
\begin{align}
    C_{\text{intra}}^\prime(\boldsymbol{q}) = \frac{1}{N^2} \sum_{i,j} \left( \langle a_i a_j \rangle - \langle a \rangle^2 \right) e^{-i\boldsymbol{q} \cdot (\boldsymbol{r_i} - \boldsymbol{r_j})},
\end{align}
where $a_i$ represents the electron number or the $z$-component of the spin at site $i$, $\langle \cdot \rangle$ denotes the expectation value of the physical quantity, $\boldsymbol{q}$ is the wave vector, and $\boldsymbol{r_i}$ is the position of site $i$. This definition of $C_{\text{intra}}^\prime(\boldsymbol{q})$ is commonly used when analyzing correlations in many-body systems.

In the main text, we introduced a modified definition of the intra-layer correlation function, $C_{\text{intra}}(\boldsymbol{q})$, which includes the restriction $i \ne j$ in the summation. Importantly, this additional constraint does not alter the peak structure of the correlation function. To demonstrate this, we consider the difference between the two definitions:
\begin{align}
    C_{\text{intra}}^\prime(\boldsymbol{q}) - C_{\text{intra}}(\boldsymbol{q}) = \frac{1}{N^2} \sum_i \left( \langle a_i a_i \rangle - \langle a \rangle^2 \right).
\end{align}
This term is independent of $\boldsymbol{q}$, implying that $C_{\text{intra}}^\prime(\boldsymbol{q})$ is simply a constant shift of $C_{\text{intra}}(\boldsymbol{q})$. Consequently, both definitions exhibit identical peak structures.

\begin{figure}
    \centering
    \includegraphics[width=\linewidth]{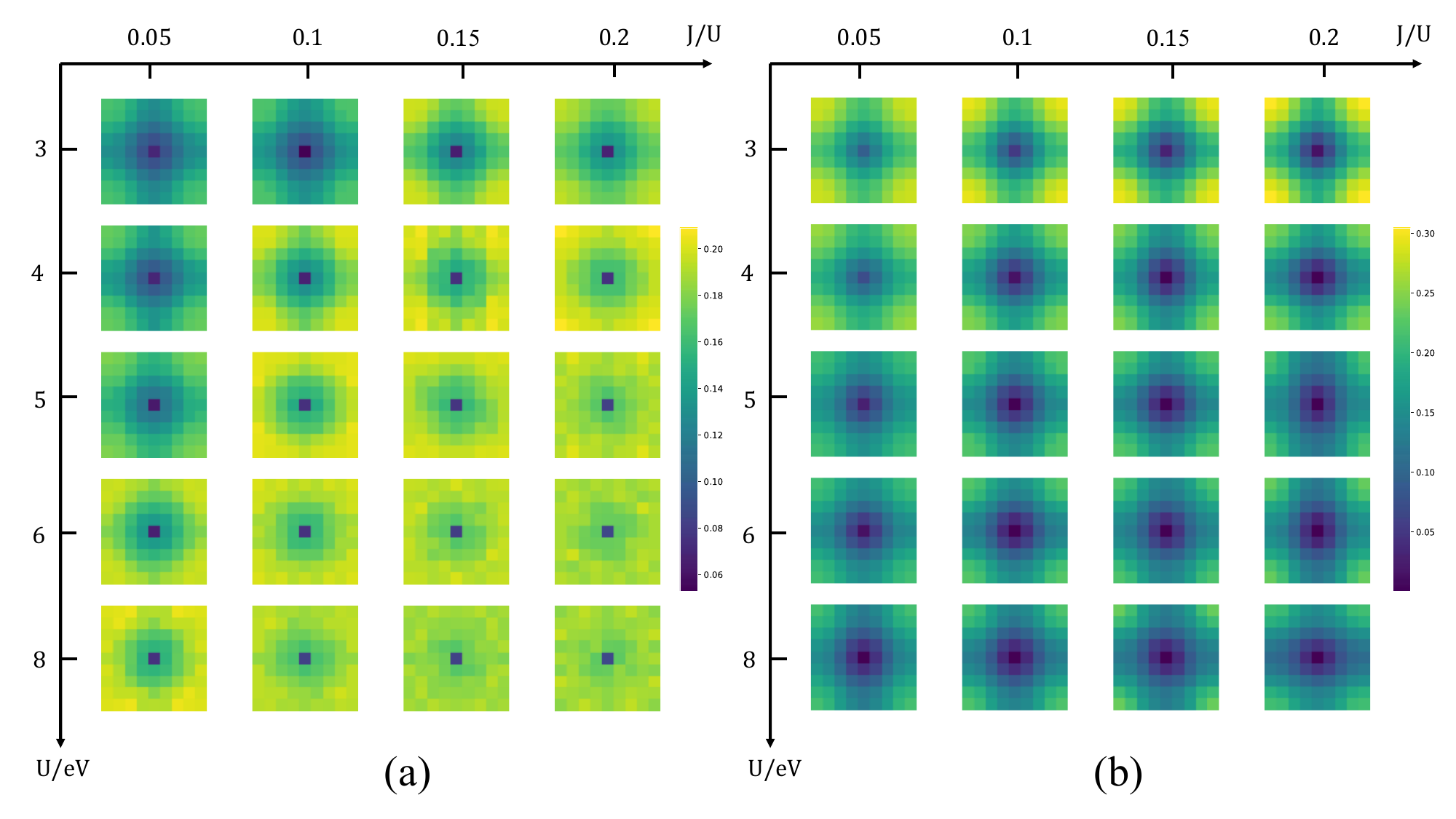}
    \caption{Results for $C_{\text{intra}}^\prime(\boldsymbol{q})$ in both spin and charge correlation functions. (a) The spin correlation function in momentum space for layer B. (b) The charge correlation function in momentum space for layer B. Both plots exhibit peaks at $(\pi, \pi)$, consistent with the results in the main text.}
    \label{SFig3}
\end{figure}

To further validate this result, we present the calculated $C_{\text{intra}}^\prime(\boldsymbol{q})$ in Fig. \ref{SFig3}. The figure shows the spin correlation function and the charge correlation function in momentum space for layer B. The peaks are consistently located at $(\pi, \pi)$, with no additional peaks observed, in agreement with the results presented in the main text.

\bibliography{ref}